\documentclass[prb,superscriptaddress,aps, bibliography, twocolumn, reprint, showpacs, footnoteinbib]{revtex4-1}

\usepackage{graphicx}
\usepackage{amssymb}   
\usepackage{amsfonts}
\usepackage{amsmath}
\usepackage{dsfont}
\usepackage{natbib}
\usepackage{dcolumn}   
\usepackage{bm}        
\usepackage[mathscr]{eucal}
\usepackage[dvipsnames]{xcolor}
\usepackage[colorlinks, linkcolor=OrangeRed,citecolor=RoyalBlue,urlcolor=NavyBlue]{hyperref}
\usepackage[all]{hypcap} 
\usepackage{mathtools}
\usepackage{dsfont}

\definecolor{myblue}{rgb}{0,0,0.75}

\newcommand{\ket}[1]{| #1 \rangle}
\usepackage{graphicx}
\usepackage{amsmath,amssymb,bm}
\usepackage{enumerate}
\usepackage{braket}        

\begin{document}

\title{Dynamics of strongly interacting systems: From Fock-space fragmentation to Many-Body Localization}
\author{Giuseppe De Tomasi}
\affiliation{Department of Physics, Technische Universit\"at M\"unchen, 85747 Garching, Germany}
\author{Daniel Hetterich}
\affiliation{Department of Physics, Technische Universit\"at M\"unchen, 85747 Garching, Germany}
\author{Pablo Sala}
\affiliation{Department of Physics, Technische Universit\"at M\"unchen, 85747 Garching, Germany}
\affiliation{Munich Center for Quantum Science and Technology (MCQST), Schellingstr. 4, D-80799 M\"unchen, Germany}
\author{Frank Pollmann}
\affiliation{Department of Physics, Technische Universit\"at M\"unchen, 85747 Garching, Germany}
\affiliation{Munich Center for Quantum Science and Technology (MCQST), Schellingstr. 4, D-80799 M\"unchen, Germany}

\begin{abstract}
We study the $t{-}V$ disordered spinless fermionic chain in the strong coupling regime, \mbox{$t/V\rightarrow 0$}. Strong interactions highly hinder the dynamics of the model, fragmenting its Hilbert space into exponentially many blocks in system size. Macroscopically, these blocks can be characterized by the number of new degrees of freedom, which we refer to as movers. We focus on two limiting cases: blocks with only one mover and the ones with a finite density of movers. The former many-particle block can be exactly mapped to a single-particle Anderson model with correlated disorder in one dimension. As a result,  these eigenstates are always localized for any finite amount of disorder. The blocks with a finite density of movers, on the other side, show an MBL transition that is tuned by the disorder strength. Moreover, we provide numerical evidence that its ergodic phase is diffusive at weak disorder. Approaching the MBL transition, we observe sub-diffusive dynamics at finite time scales and find indications that this might be only a transient behavior before crossing over to diffusion.  
\end{abstract}
\maketitle
\section{Introduction}
Recent advances in controlled experimental techniques on ultracold atoms~\cite{Bloch08,Bloch2012} in optical lattices and trapped ions~\cite{Georg14,Blatt2012} allow to inspect dynamical properties
of closed quantum disordered systems and to provide signatures for the existence of a many-body localized phase~\cite{Bloch2015,schneider2015,choi2016exploring,monroe2015,Luschen17} (MBL). An MBL phase describes a perfect insulator in which interacting particles are localized due
to the presence of a strong disordered potential, generalizing the phenomenon of Anderson localization~\cite{Evers2008Review, Anderson1958} to the many-body case~\cite{Basko06,gornyi2005interacting,nandkishore2015many}.
Moreover, the non-interacting localized eigenstates are adiabatically connected to the MBL eigenstates~\cite{ros2015integrals, imbrie2017review, serbyn2013local, Abanin2016Explicit, huse2013phenomenology,GDT19, imbrie2014many}, which implies that an MBL-phase is fully described by an extensive number of quasi-local integrals of motion, which emphasize an emerging weak form of integrability~\cite{ros2015integrals, imbrie2017review, serbyn2013local, Abanin2016Explicit,huse2013phenomenology, GDT19,imbrie2014many}.

The MBL phase should be opposed to the ergodic one, in which local observables reach their thermal equilibrium, eigenstates are believed to be chaotic and dynamics shows delocalization~\cite{Bera17,Luitz15, luitz2017ergodic, Luitz16, La16, Ser15,Bera15, Giu17,Pro08,Pal10}. 
The ergodic phase is characterized by the eigenstate thermalization hypothesis~\cite{deutsch1991quantum, Srednicki1994, Srednicki1996,RigolOlshanii, Eisert2016Rev} (ETH), which asserts that the system locally thermalizes at the level of single eigenstates.  

Recently, it has been shown experimentally and numerically that the two aforementioned scenarios are not the only possibilities. For example, the relaxation of kinetically constrained many-body systems (e.g. Rydberg-blockaded chains) could be extremely slow, if prepared in specific experimentally accessible out-of-equilibrium initial states~\cite{ Lukin17,  Tur18, KhemaniV19, Turner2018,Choi19, Ho19}. 
This reminiscence of integrable behavior for the dynamics of certain initial states is believed to be captured by a set of eigenstates of measure zero, which violates ETH~\cite{ Turner2018, Mou18,KhemaniV19, Ho19} and have a considerable big overlap with these initial states. Importantly, these \textit{atypical} eigenstates  are even at infinite temperature distributed through the whole spectrum, such that they are embedded into a sea of thermal states.    
These atypical eigenstates may remind to the concept of quantum scars, a measure zero sets of quantum eigenfunctions localized around unstable classical periodic orbits in quantized chaotic systems~\cite{Antosen95, Heller84, KAPLAN1998171}, e.g. quantum biliards~\cite{Haake06}. 

Although it is still under debate whether these states can be considered as a many-body generalization of quantum scars, they have a peculiar characteristic: 
they live in a small portion of the Fock-space, which usually scales only polynomially with system size. As a consequence these eigenstates have a highly non-thermal behavior (e.g. low entanglement).

The same effect is obtained if the \textit{entire} Fock-space splits into different blocks. This was recently described in Refs.~\onlinecite{Sala19,Khemani19} as a  \textit{Hilbert space fragmentation}, where a system is hindered to thermalize due to dynamical constraints, which separates the Fock-space into exponentially many disjoint invariant subspaces. 
 Due to this fragmentation, the system exhibits non-thermal eigenstates appearing throughout the entire spectrum, thus breaking the strong formulation of ETH.  Similar ideas have been recently used to provide a decimation scheme to study the MBL transition in the random field Heisenberg model in Ref.~\onlinecite{Pietracaprina19}.

These paradigms of ergodicity breaking brought new emphasis and stimulated extensive research attempting to understand the thermalization properties of quantum matter. The aim of this work is to shed light on the nature and origin of the just mentioned phenomenology in the presence of a disordered potential. To this end, we study 
the $t{-}V$ disordered spinless fermionic chain, which is believed of having an MBL transition, in the  strong coupling limit ($t/V \rightarrow 0$). In this regime the model is equivalent to a dynamically constrained model. Moreover, due to the presence of strong constraints, its Fock-space fragments into exponentially many blocks. Macroscopically these blocks can be characterized by the number of some  degree of  freedom, that we call \emph{movers}, which are responsible of the remaining dynamical properties. 

We study the out-of-equilibrium dynamics focusing  on two complementary limiting cases: blocks with only one mover and blocks with a finite density of movers. The former can be mapped to a single-particle Anderson localization problem on the Fock-space with correlated disorder. As a result, the model restricted on these blocks with zero density of movers, is always localized for any finite amount of disorder. Instead, for the blocks with a finite density of movers, we provide evidence for an MBL transition between states that can be considered thermal within the block and localized ones. 

Importantly, due to the constrained dynamics, less disorder is required in order to localize the model than in the case of finite interaction strength $V$. Furthermore, we study the dynamics on the ergodic side of the blocks with finite density of movers. Here, our results are consistent with the existence of diffusive dynamics at weak disorder. At stronger disorder, approaching the MBL transition, we see a clear slow long-time crossover from a transient sub-diffusive dynamics to a diffusive one. As expected, the  time  scale for  the  onset  of  the diffusive propagation shifts to infinity on approaching the MBL transition~\cite{Basko06, gornyi2005interacting, Bera17}. Thus, we provide evidence for an extensive region within the ergodic phase characterized by a small diffusion constant, which could be identified with the ``bad-metal'' phase predicted by Basko, Aleiner and Altshuler in their seminal work~\cite{Basko06}.  

The  rest  of  the  work is  organized  as  follows. In Sec.~\ref{sec:Model} we introduce the model and we discuss its strong interaction limit. In Sec.~\ref{sec:Frag} we explain in detail the constrained dynamics on the Fock-space. In Sec.~\ref{few_movers} we inspect the dynamical properties of the block with one mover. We map the model to an Anderson model with correlated disorder. Here, we comment on the relation between blocks with few movers and many-body scars. Finally, in Sec.~\ref{sec:Finite_movers} we study both eigenstates and dynamical properties of the largest block of the Hamiltonian (finite-density of movers). The Sec.~\ref{sec:Finite_movers1} is dedicated
to show the existence of the MBL transition by studying both spectral and eigenstates properties. Using finite-scaling techniques we analyze several quantities (i.e. energy level statistics, entanglement) and we extract an estimation for the critical point. In Sec.~\ref{sec:Finite_movers2} we focus on the dynamical properties inspecting the relaxation of the density propagator. In this section we provide numerical evidence that its ergodic phase is diffusive. In Appendix we extend our work to the case in which the disorder is generated by a quasi-periodic potential and to the case of large but finite interaction strength. 

\begin{figure}
\includegraphics[width=1.\columnwidth]{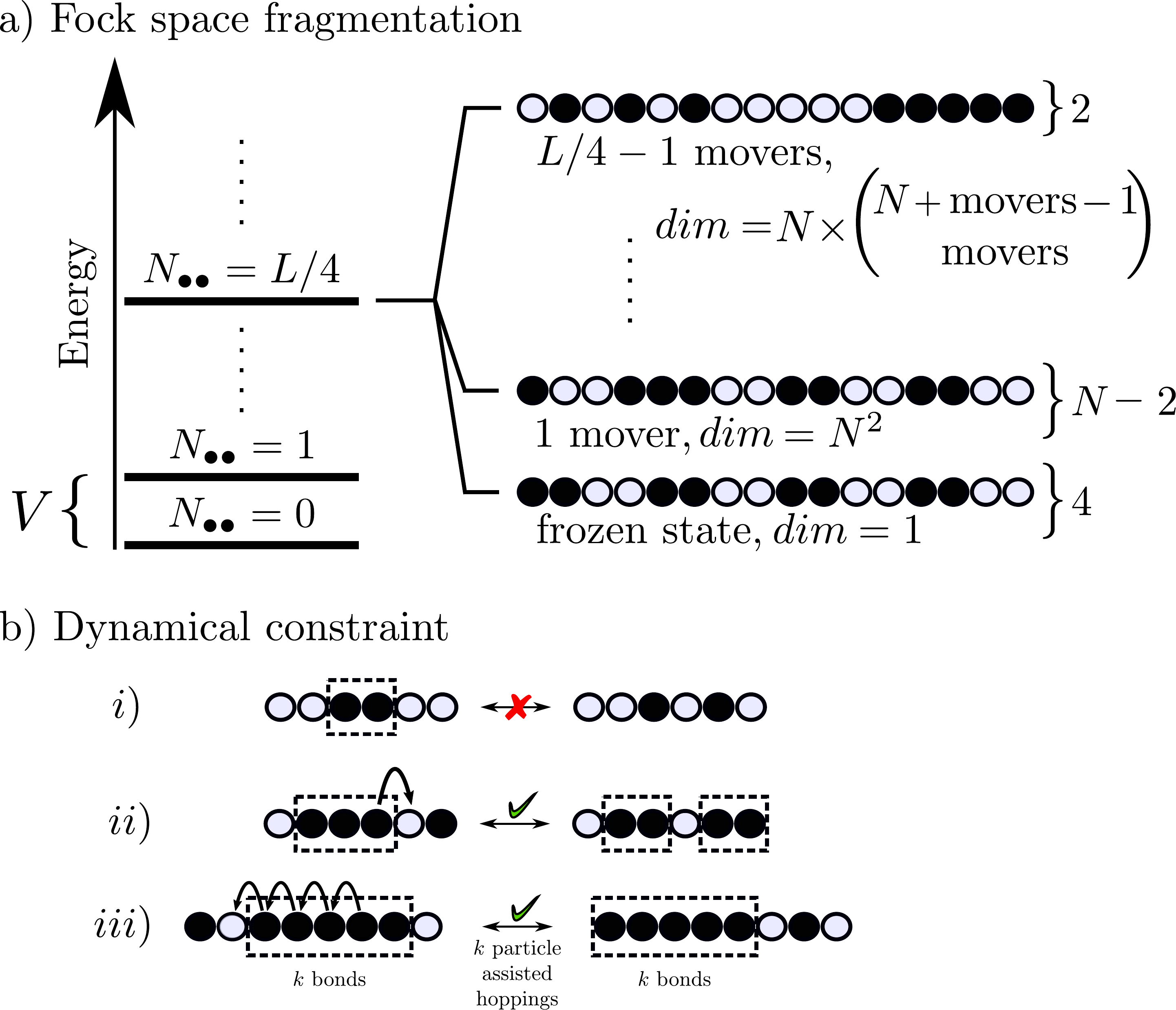}
\caption{(a): Illustration of the disconnected sectors of the Hamiltonian $\hat H_{\infty}$ in Eq.~\ref{eq:H_inf}. For fixed particle number $N=L/2$, the sectors can be distinguished by the number of bonds $N_{\bullet\bullet}$, i.e. the number of adjacent particles. Sectors with different values of $N_{\bullet\bullet}$ are energetically separated by an energy scale of order $\sim V$. For fixed $N$ and $N_{\bullet\bullet}$, there exist an exponential number in $L$ of disjoint sectors (see text). They can be further classified by the number of delocalized degree of freedom, which we call movers.
(b): The limit $t/V\to 0$ imposes constraints on the dynamics of particles. $i)$ A block of particles cannot split on its own as the number of bonds $N_{\bullet\bullet}$ is conserved. A change of this number would imply an energy cost $\sim V$, assumed to be large. $ii)$ Particles at the edge of a block can leave the block if a second particle assists the hop, such that $N_{\bullet\bullet}$ stays constant. $iii)$ This mechanism allows for an effective tunneling of a mover through a block, where the block always moves two sites towards the opposite direction (assisted hopping). }
\label{fig:rules}
\end{figure}

\section{Model} \label{sec:Model}
We study the $t{-}V$ disordered spinless fermionic chain with periodic boundary conditions,

\begin{equation}
\label{eq:full_H}
 \hat H = -t \sum_x \hat c_{x+1}^\dagger \hat c_{x} + \textit{h.c.} + W\sum_x \mu_x \hat n_x + V\sum_x \hat n_x \hat n_{x+1},
\end{equation}
where $\hat c^\dagger_x$ ($\hat c_x$) is the fermionic creation (annihilation) operator at site $x$, $\hat n_x = \hat c_x^\dagger \hat c_x$ and $\mu_x$ are independent  random  variables  uniformly  distributed in $[-1,1]$. In the following we fix time and energy scales with respect to the hopping amplitude $t=1/2$, and study the behavior of the system as a function of $W$ and $V$, the disorder and interaction strengths respectively. Moreover, $L$ is the number of sites and we restrict our analysis to half-filling, i.e. the number of particles $N=L/2$.

The non-interacting limit $V=0$ is the Anderson model and all its single-particle wave functions are exponentially localized for any finite amount of disorder~\cite{Evers2008Review, Anderson1958, mott1961theory}. At finite interaction strength, this model is believed to have an MBL transition for 
$V=1$ at $W_c \approx 3.5$~\cite{Luitz15, Soumya2015SingleParticlePerspective, Giu17, Serbyn2015criterion, serbyn2014quenches} ($W<W_c$ ergodic and $W>W_c$ localized). 
In this work, we instead focus on large interaction strengths, i.e. $V \gg t,W$. As a first approximation, we consider the limit $V/t\to\infty$~\cite{Dias2000,Peres99,Sutherland90,hot_17,Bar_lev_16_prb}. In this regime the spectrum of $\hat{H}$ splits into energetically separated bands composed by states with identical number of pairs of nearest-neighbor occupied sites $N_{\bullet\bullet} \equiv \sum_x \hat n_x \hat n_{x+1}$, which we name \emph{bonds}~\cite{Dias2000}. Projecting $\hat H$ into each of these bands we obtain the following effective Hamiltonian to first order in perturbation theory~\cite{Dias2000,Peres99} 
\begin{equation}
\label{eq:H_inf}
 \hat H_{\infty} =   -t\sum_x  \hat P_{x} \big( \hat c_{x+1}^\dagger \hat c_{x} + \textit{h.c.} \big)\hat P_{x}  + W\sum_x \mu_x \hat n_x,
\end{equation}
with the local projector
\begin{equation}
 \hat P_{x} = 1 - (\hat n_{x+2} - \hat n_{x-1})^2, \qquad \hat P_{x}^2 = \hat P_{x},
\end{equation}
that guarantees that a particle can only hop if the number of bonds $N_{\bullet\bullet}$ is unchanged. 
 Hence, by construction, the number of bonds $N_{\bullet\bullet}$ is conserved ($[N_{\bullet \bullet}, \hat H_{\infty}]=0$), which strongly constrains the dynamics of the model, as expressed by the presence of the local projectors $\{\hat P_{x}\}$ in the kinetic terms \footnote{In fact, the Hamiltonian $\hat H_{\infty}$ takes the form described by Siraishi and Mori in Ref.~\onlinecite{Siraishi17} for which one can show the existence of ETH violating eigenstates.}. In the remain of the work, we focus on the largest band $N_{\bullet \bullet} = L/4$, which is in the middle of the spectrum of $\hat H_{\infty}$ (see Fig.~\ref{fig:rules}~(a)). The dimension of this symmetry subspace $\mathcal{H}$  with $N=L/2$ and $N_{\bullet \bullet}=L/4$  is given by $\text{dim} (\mathcal{H})= \binom{L/2}{L/4}^2\sim \frac{2^L}{L}$, thus up to polynomial corrections in $L$ it covers the full Hilbert  space of $\hat H$ (Eq.~\ref{eq:full_H}).  

The conservation of $N_{\bullet\bullet}$ strongly restricts the dynamical features of our model, because the creation or annihilation of a bond would cost an infinite amount of energy $V$. Hence, a particle can only jump from a site $x$ to the neighboring site $x+1$  if both adjacent sites $x-1$ and $x+2$  are either simultaneously occupied or empty. 

The dynamics can be described as following: 
separated single particles, only surrounded by holes, are free to move. Domains of particles instead are stuck because the number of bonds is conserved. Importantly, however, if a separate particle approaches a block, it can assist a hop of a domain particle. Following this mechanism step by step, a particle and a domain can interchange their positions, whereby the domain effectively moves as a whole by two sites into the direction where the particle originally was. These rules for the possible hops of particles are illustrated in Fig.~\ref{fig:rules}~(b) and more precisely described in the following.

We note that in the absence of on-site potential, $W=0$, the Hamiltonian $\hat H_\infty$ in Eq.~\ref{eq:H_inf} is integrable and it can be exactly solved using Bethe ansatz techniques~\cite{Peres99,Sutherland90}.
Alternatively, $\hat H_{\infty}$ can be mapped to spin $1/2$ degrees of freedom on the bonds~\cite{Dias2000}. As explained in Ref.~\onlinecite{Dias2000}, two consecutive filled $|\bullet \bullet\rangle$ (empty $|\circ\circ\rangle$) sites are identified with a spin up  $|\uparrow\rangle$ (down $|\downarrow\rangle$) on the middle bond and a mobile particle $|\circ \bullet\rangle$ is mapped to an empty bond $|\bm{0} \rangle$.

For the sake of completeness and to better explain this mapping, let us consider a concrete example with $L=12$ restricted to the symmetry sector of our interest, i.e.,  $N=6$ and $N_{\bullet\bullet} = 3$. 

Due to the global conserved quantities, $N=L/2$ and $N_{\bullet \bullet}=L/4$, we will have that the number of spins up $N_{\uparrow}$ is equal to the number of spins down $N_{\downarrow}$ ($N_\uparrow+ N_\downarrow = 2 N_{\uparrow} = N$) which are both independently conserved. In fact, the mobility restrictions of the spinless fermions due to the Hamiltonian $\hat H_{\infty}$, gives rise to the conservation of the spin configuration along the evolution. This means that the only allowed dynamics is a reshuffling of the position of the empty bonds $\ket{\bm{0}}$ while keeping the relative orientation of each spin unchanged. 

The Fock state $|\bullet \bullet \circ \circ  \bullet \bullet \circ \circ  \bullet \bullet \circ \circ\rangle  $ is mapped to the N\'{e}el state in the spin configuration $| \uparrow \downarrow  \uparrow  \downarrow \uparrow \downarrow \rangle$, which is characterized by the absence of the empty bond $|\bm{0}\rangle$.

Introducing a defect on the former pattern $|\bullet \bullet \circ \circ \bullet \bullet \bullet \circ\circ  \bullet \circ\circ   \rangle$, we obtain the following spin configuration $| \uparrow\downarrow\uparrow \uparrow \downarrow \bm{0} \downarrow\rangle$ which has one empty bond $|\bm{0}\rangle$. This empty bond $|\bm{0}\rangle$ is now free to hop $|\bullet \bullet \circ \circ \bullet \bullet \bullet \circ\circ  \bullet \circ\circ   \rangle \rightarrow |\bullet \bullet \circ \circ \bullet \bullet \bullet \circ\bullet \circ \circ \circ   \rangle \Rightarrow | \uparrow\downarrow\uparrow \uparrow \downarrow \bm{0} \downarrow\rangle \rightarrow | \uparrow\downarrow\uparrow \uparrow  \bm{0} \downarrow \downarrow \rangle$, while keeping the same spin configuration. Thus, empty bonds $|\bm{0}\rangle$ are the delocalized degrees of freedom moving around the spin pattern. The pattern remains fixed during the dynamics of the empty bonds $|\bm{0}\rangle$ up to cyclic rotations and thus the number of flips $|\downarrow \uparrow \rangle$ in the spin configuration $N_{\downarrow \uparrow}$ is a constant of motion. 

For the remainder of the work, we name the number of empty bonds $|\bm{0}\rangle$ \textit{movers}. However, in general, the number of movers is not a constant of motion due to a subtle detail in the definition of the spin mapping~\cite{Dias2000}.  Therefore, to simplify this discussion we introduce an additional rule: when a mover approaches from the left a domain wall constituted by frozen particles, e.g. $|\cdots\circ\bullet\circ \circ\bullet\bullet\circ\circ \cdots \rangle \rightarrow |\cdots \circ\circ\bullet \circ\bullet\bullet\circ\circ \cdots\rangle$, and a new empty bond is generated $| \cdots \bm{0} \downarrow  \uparrow \cdots\rangle \rightarrow | \cdots \downarrow \bm{0}\bm{0} \uparrow \cdots\rangle  $, this should be counted as a single empty bond ($| \cdots \downarrow \bm{0} \uparrow \cdots\rangle$). These processes describe an assisted hopping, in which a free particle activates a particle belonging to a domain helping it to escape. With this definition the number of movers $N_{\bm{0}}$ is independently conserved allowing us to further characterize the connected sectors. As further consequence, the sum between the number of antiferromagnetic aligned neighbor spins  $|\downarrow \uparrow \rangle$ ($N_{\downarrow\uparrow}$) and the number of movers $N_{\bm{0}}$ is constant with the important relation $N_{\downarrow \uparrow} + N_{\bm{0}} = N_{\bullet\bullet}$.



 Using this mapping, $\hat{H}_\infty(W=0)$ can be rewritten as a strongly coupled Hubbard chain that is subject to a fictitious flux proportional to the total momentum~\cite{Dias2000}. Moreover, the optical conductivity of $\hat H_\infty(W=0)$ at low temperature was calculated showing that it is a perfect insulator~\cite{Peres99,Zotos96,Castella95,Zotos97}.  

Summarizing, the constrained  dynamics induced by $\hat H_\infty$ of an initial state in the Fock-space is determined by the number of movers and by the spin configuration, in the way  just described. As a result, due to the exponential number of spin configurations within the same global symmetry sector, the Fock-space of $\hat H_\infty$ fragments into exponentially many blocks in system size as we will describe in the next section.

\section{Fragmentatiom of the Fock-space} \label{sec:Frag}

Although the full characterization of disjoint blocks is out of the scope of this work, in this section we will discuss some of the block structure of $\hat H_\infty$ in Eq.~\ref{eq:H_inf}    . 

\begin{figure}
\includegraphics[width=1.\columnwidth]{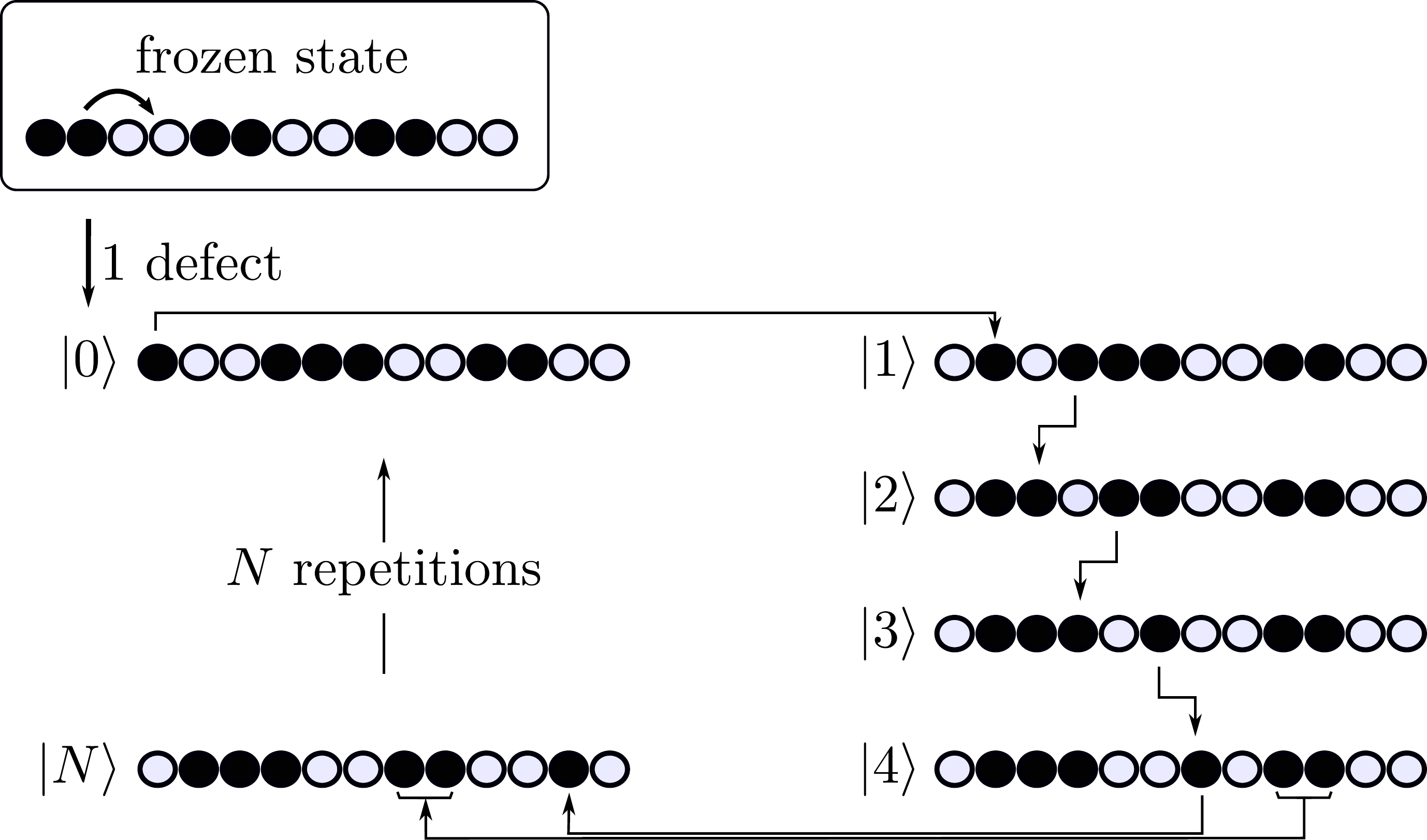}
\caption{(a): Illustration of the $N^2$ dimensional subspace that describes a single moving particle.
An empty site is created from the frozen state by moving one particle to an adjacent block of particles.
This leaves a single particle that can change position with a block while respecting the global conservation laws of the Hamiltonian (see steps from $\ket{0}$ to $\ket{4}$. After $N$ steps, the particle crossed all present blocks and, effectively, each particle moved two sites to the left (compare $\ket{0}$ with $\ket{N}$). Thus, repeating this procedure $N$ times, each particle moved $2N=L$ sites to the left, such that the state $\ket{0}$ is again obtained. The full space thus contains $N^2$ states. Moreover, the states are connected like a one-dimensional chain of $N^2$ sites.}
\label{fig:1defect}
\end{figure}

First, it is easy to see that there exist four frozen states that are disconnected from the remaining symmetry subspace and therefore show no dynamics. They are obtained by an iterative filling of the chain with two adjacent particles and two adjacent holes (e.g.  $|\bullet \bullet \circ\circ  \bullet \bullet \circ\circ  \cdots\rangle $). Such a state is shown at the top of Fig.~\ref{fig:1defect}.  The other frozen states are obtained by cyclic rotations of this state.  In the spin language using the map described in the previous section~\cite{Dias2000}, the frozen states are the N\'{e}el states, e.g. $|\uparrow \downarrow  \uparrow \downarrow  \uparrow \downarrow   \cdots \rangle$,
without movers $|\bm{0}\rangle$, i.e. $N_{\downarrow\uparrow}=N_{\bullet\bullet}$. 

Creating a defect on one of these, i.e., releasing a particle from one of the existing bonds, without changing the number of bonds, as shown in Fig.~\ref{fig:1defect}, a new state is formed that consists of a \emph{mover}, a bond domain with three particles and all original bonds of the frozen state 
(e.g. $|\bullet \circ\circ \bullet\bullet \bullet \circ\circ \bullet \bullet \circ\circ \rangle$). Using the mapping described in the previous section, this state is mapped to a state with one mover (empty bond $|\bm{0}\rangle$) and $N_{\bullet \bullet}-1$ spin flips $|\downarrow \uparrow\rangle$ (e.g. $|\bm{0}\downarrow \uparrow \uparrow \downarrow \uparrow \downarrow\rangle$). It is important to remember that the number of spin up is equal to number of spin down ($N_{\uparrow} = N_{\downarrow} = N/2 = N_{\bullet \bullet}$), and the sum of number of movers and spin flips is equal to $N_{\bm{0}} + N_{\downarrow \uparrow} = N_{\bullet \bullet}$. Since the number of movers and the spin configurations up to rotations are invariant of motion, this block is described by one mover and a configuration of $N$ spins with zero magnetization ($N_{\uparrow} = N_{\downarrow}$) and $N_{\bullet\bullet} -1$ spin flips ($N_{\downarrow\uparrow}$). 

Following the rules of Fig.~\ref{fig:rules}, the mover is able to hop around the chain and assist other particles to translate each domain wall by two lattice sites, see Fig.~\ref{fig:1defect}. Using the periodic boundary conditions and repeating this scheme as shown in Fig.~\ref{fig:1defect}, one sees that the invariant subspace containing this specific one mover state has dimension $N^2 =(L/2)^2$. 
Indeed, in the spin language, the mover hops around the fixed spin configuration which is shifted by one site once the mover crosses the whole spin chain. As a consequence, we will have $N$ places where the mover can be times $N$ possible rotations~\footnote{This prefactor $N$ dependence of the cyclic symmetry of the spin configuration} of the spin configuration, giving thus the $N^2= N\cdot N$ dimension of the block.  

Moreover, these $N^2$ states are connected to each other in a ring-like manner, see Fig.~\ref{fig:rules}. This is why we can map the blocks consisting of one mover to a model where a single particle moves in a one-dimensional chain with $N^2$ lattice sites. 

It is important to realize that even for a fixed number of movers, many disconnected blocks exist. As discussed, for zero movers, i.e. frozen states, four such blocks (of dimension 1) exist. The above constructed $N^2$-dimensional block that describes a single mover is degenerate $N-2$ times. This can be seen as follows: The construction of the single mover shown in Fig.~\ref{fig:1defect} creates a domain of three particles, which is separated by three empty sites from the next domain to the left side, if the mover is not directly between these two domains. This can be seen in state $\ket{3}$ or $\ket{4}$ in Fig.~\ref{fig:1defect}. However, we could have constructed a single mover in a way such that the three consecutive empty sites are between any of the $N/2-1$ domains. In each of these setups, we can further shift all particles by one site, which also yields a new disconnected block, because, as discussed above, domains may only move in steps of 2 sites if a mover crosses them. Thus there exist $N-2$ blocks of dimension $N^2$ that describe a single mover. 

The maximum number of movers that the system can host is $N_{\bullet \bullet}-1$, which as expected gives rise to the largest connected sectors within the global symmetry subspace. Using the map to spins it is equivalent to $N_{\bullet \bullet}-1$ empty bonds ($N_{\bm{0}} = N_{\bullet \bullet}-1$) and one spin flip $(N_{\downarrow\uparrow} = 1$), thus fulling the global constraint $N_{\bm{0}} + N_{\downarrow \uparrow} = N_{\bullet \bullet}$. The resulting spin configuration is the domain wall (e.g. $|\downarrow\downarrow\downarrow\uparrow\uparrow\uparrow \rangle$) up to cyclic rotations. The dimension of a block with $N_{\bullet\bullet}-1$ movers is $N \binom{N + N_{\bullet \bullet}-2}{N_{\bullet \bullet}-1}$. The combinatorial factor $\binom{N + N_{\bullet \bullet}-2}{N_{\bullet \bullet}-1}$ is the way that one can dispose $N_{\bm{0}}=N_{\bullet \bullet} -1$ movers within the domain spin configuration and the factor $N$ comes from the cyclic property of the latter~\cite{Dias2000}.

In general, $\hat H_\infty$ in Eq.~\ref{eq:H_inf} has a block structure, in which disjoint blocks are characterized by the number of movers and the spin configuration (up to cyclic rotation) fulling the global constrains $N_\uparrow = N_\downarrow = N/2$  and $N_{\bm{0}} + N_{\downarrow\uparrow} = N_{\bullet\bullet}$.  The dimension of each block is given by $g  \binom{N + N_{\bm{0}}-1}{N_{\bm{0}}}$, where $g\cdot \in \mathbb{N}$ is an integer number $g\le N$ that counts the number of different configurations obtained by cyclic rotations. Moreover, due to the exponential number of spin configurations, the system has at least $N^{-1} \binom{N}{N_{\bullet\bullet}}= (N_\uparrow + N_\downarrow)^{-1} \binom{N_\uparrow + N_\downarrow}{N_\uparrow }$ disjoint blocks. 

In Appendix we will give an alternative argument for the exponential fragmentation of the Fock-space, which is not based on the spin mapping. 

In the following sections we study the dynamics of the system in two limiting cases. First, the case in which only one mover is present. We will map the system to a single-particle localization problem on the Fock-space. Consequently, the system will be exponentially localized for any amount of disorder. Second, we study the system with a finite  density of movers.  The latter case shows an MBL transition between ergodic and localized states. Moreover, we will provide evidence that the dynamics of the system might be diffusive on its ergodic side.

\section{Many-body scars} \label{few_movers}

In this section we inspect the dynamical properties of $\hat{H}_\infty$ within the blocks with a single mover.   Considering the $N^2=(L/2)^2$ many-body Fock states $\ket{j}$ contained in such blocks as an effective single-particle basis, as it is shown in Fig.~\ref{fig:1defect}, $\hat H_\infty$ can be mapped to a one-dimensional chain
\begin{equation}
\label{eq:map}
\hat H_\infty = -\frac{1}{2}\sum_j | j\rangle \langle j+1 | + \text{h.c.} + \sum_j \chi_j |j\rangle \langle j|,
\end{equation}
of $N^2$ lattice sites $j$, where $\chi_j = W \sum_x \mu_x \langle j | \hat n_x | j \rangle$ is the effective potential on the site-state $|j\rangle$. 

Our main observable is the return probability starting from a Fock state $|j\rangle$
\begin{equation}
\label{eq:return_prob}
 R(t) = |\langle j | e^{-i\hat H_{\infty} t} |j \rangle |^2,
\end{equation}
where we average over disorder and different initial states within the block, which from now on will be indicated with an overline, i.e. $\overline{R}(t)$. Such initial states are experimentally accessible in cold-atom set up experiments as they are product states in the local particle number basis. 

Recently, the return probability  $R(t)$  in Eq.~\ref{eq:return_prob} has been studied in kinematically constrained models like Rydberg-blockaded chains, showing that when the system is initialized in a specific initial state, $R(t)$ shows slow dynamics, even though the system is believed to be thermal~\cite{KhemaniV19, Turner2018, Mou18, Ho19}. It has been argued that the reason for this slow dynamics is a set of eigenstates with measure zero in the full Hilbert space, called many-body scars~\cite{KhemaniV19, Turner2018, Mou18, Ho19}, which have a considerable high overlap with the initial state. One of the main characteristic of these many-body scars is the fact that are highly non-thermal since they live in a small portion of the Fock-space.

For sake of completeness, we start our investigation from the free-disorder case, $W=0$. In this case our model equals to a free hopping problem on the fictitious one-dimensional chain of $N^2$ sites. Thus, 
\begin{equation}
\label{eq:exact_W_0}
 R(t) = \mathcal{J}_0^2(t),
\end{equation}
where $\mathcal{J}_0(t)$ is the Bessel function of the first kind ($\mathcal{J}_0^2(t) \approx \frac{2}{\pi t} \cos^2 (t+ \pi/4)$). 

\begin{figure}
\includegraphics[width=1.\columnwidth]{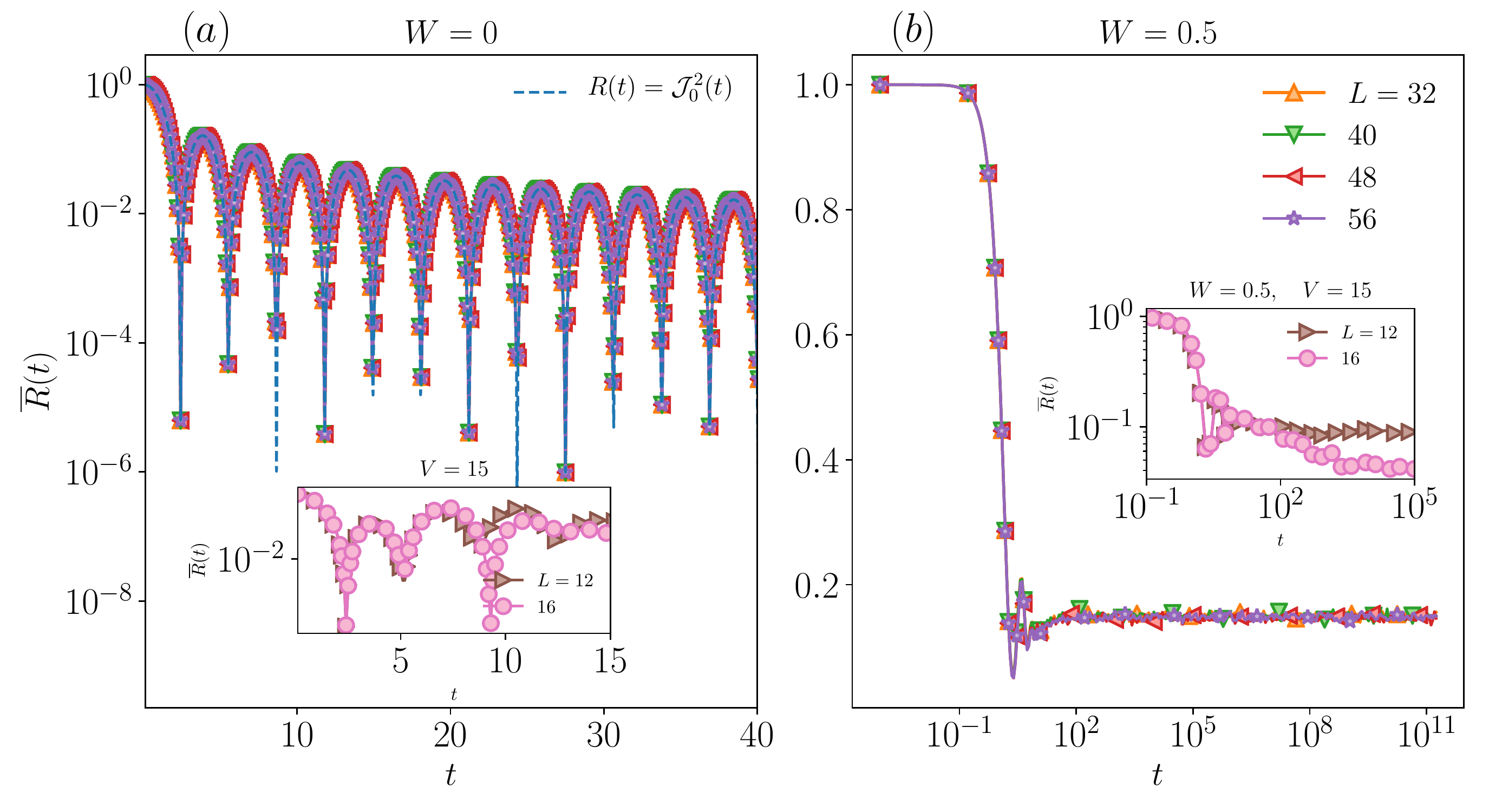}
\caption{(a): $R(t)$ for the disorder-free case $W=0$ for several system sizes $L\in \{32,40,48,52\}$. The dashed line is the theoretical prediction $R(t) = \mathcal{J}_0^2(t)$. The inset shows $R(t)$ computed using the Hamiltonian $\hat{H}$ with $V=15$. (b): $R(t)$ for disorder strength $W=0.5$ and $t/V\rightarrow 0$. $R(t)$ saturates with time to an $L$ independent value, meaning that the system is localized. The inset shows $R(t)$ computed with $\hat H$ in Eq.~\ref{eq:full_H} for $V=15$. Here, $R(t)$ starts to decay at time scale of order $\sim V^2$, at which the next order corrections ($t/V$) become relevant. }
\label{fig:Fig1}
\end{figure}

Figure~\ref{fig:Fig1}~(a) shows the return probability $R(t)$ (Eq.~\ref{eq:return_prob}) for the free disorder case ($W=0$) computed with exact diagonalization and the exact solution in Eq.~\ref{eq:exact_W_0} (dashed line in Fig.~\ref{fig:Fig1}~(a)). We have also controlled the robustness of our perturbative approach ($t/V \rightarrow 0$), by calculating $R(t) = |\langle j | e^{-i\hat H t} |j \rangle |^2$ using the full Hamiltonian $\hat H$ at  strong interaction strength $V=15$, as shown in the inset of Fig.~\ref{fig:Fig1}~(a). Therefore, the physics of this restricted subspace is experimentally accessible in the case of large but finite interaction strength $V$ to certain time scale.

In the case $W\ne 0$, the Hamiltonian $H_\infty$ is equal to an  Anderson model with correlated on-site disorder given by $\{\chi_j\}$. We emphasize that the disorder is correlated because the fictitious potentials $\chi_j = W \sum_x^L \mu_x \langle j | \hat n_x | j \rangle$ depend on the particle configuration on the chain determinated by each Fock state $\{ |j\rangle\}$. In this way, $L$ uncorrelated uniformly distributed random fields $\{\mu_x\}$ determine the $N^2=(L/2)^2$ on-site potentials $\{\chi_j\}$ of the fictitious lattice. 
Anderson localization in one-dimension with short-range hopping is a rather stable phenomenon, which breaks down only for fine-tuned kinds of correlations in the disorder, for instance long-range correlations~\cite{Croy2011, PhysRevLett.81.3735, IZRAILEV2012125}. Indeed, one can calculate the correlation function $\overline{ (\chi_i - \overline{\chi_i})(\chi_j - \overline{\chi_j})}$ to realize that no special structure with the distance $|i-j|$ is present  and thus the system should be localized in the Fock-space. 
To test this statement, we compute the averaged return probability $\overline{R}(t)$ for a fixed weak disorder strength $W$. As shown in Fig.~\ref{fig:Fig1}~(b), the return probability  saturates to a finite and $L$-independent value at long times ($\lim_{t\rightarrow \infty} \overline{R}(t)\ne0$). Thus, the system restricted on the blocks with one mover is localized and its eigenstates are highly non-thermal.

Although we consider the limit of infinitely large interaction strength ($t/V\rightarrow 0$), we can also predict the relevant time scale until which this saturation will be observed when considering the dynamics of $\hat H$ in Eq.~\ref{eq:full_H} for finite but large $V$ ($V\gg t, W$). The inset of Fig.~\ref{fig:Fig1}~(b) shows $\overline{R}(t)$ at weak disorder $W=0.5$ and $V=15$ starting from one of the states in the considered $N^2$-block. In this regime $\hat H$ is believed to be thermal~\cite{Reich15} and one will expect a fast relaxation  for $R(t)$. However, $R(t)$ shows a slow relaxation characterized by an almost frozen dynamics at intermediate time scales, see the plateau in time in the inset of Fig.~\ref{fig:Fig1} ~(b). This pre-thermal plateau is a left-over from the fully-localized phase in the limit $t/V\to 0$ and still holds up to intermediate times in the limit $V\gg t$. Within this time, the system can be considered as being localized. At longer times $t\sim V^2$ (set by the next order in perturbation theory on $1/V$), as expected our approach is not anymore controllable and the system becomes delocalized and $R(t)$ starts to decay (behavior after the plateau in the inset of Fig.~\ref{fig:Fig1}).

In Appendix we show the return probability for a $\textit{typical}$ initial state that belongs to a block with a finite density of movers. In this case $R(t)$ shows the usual fast relaxation of thermal systems, i.e. without the pre-thermal plateau. 
Indeed, in the next section we will show that even in the limit $t/V\rightarrow 0$ the system restricted in the blocks with finite density movers are thermal at weak disorder. In the thermodynamic limit the blocks with a finite-density of movers, will dominate the behavior of $\hat H$, since their dimension is parametrically  larger than the one with few movers, as it is shown in the Appendix.

As a consequence, the behavior of $R(t)$ starting from a $\textit{typical}$ state will be thermal. However, as we have shown, there exist $\textit{atypical}$ initial states for which at time scales of the order $\sim V^2$ the system can be considered localized. As we discussed, the existence of these atypical states can be understood as their large overlap with non-thermal eigenstates which belong to the blocks with few movers. 

Thus, in the spirit of other works~\cite{ Lukin17,  Tur18, KhemaniV19, Turner2018,Choi19, Ho19}, these states could be refereed as many-body scars: $\textit{atypical}$ eigenstates, which live in a small portion of the Fock-space ($\sim N^2$) and are responsible for athermal behavior of the system if properly initialized.

\section{MBL transition and Diffusion} \label{sec:Finite_movers}

In the following, we study the ergodic-MBL phase transition in the presence of a disorder potential for sectors with a finite density of movers in the thermodynamic limit, for which we obtain spectral and dynamical evidences of an MBL transition.

\subsection{Spectral and Eigenstates properties} \label{sec:Finite_movers1}
Here, we start with both spectral and eigenstates properties of $\hat H_{\infty}$. In particular, we focus our attention on one of the largest blocks of $\hat H_\infty$, where the number of movers equals to $N_{\bullet\bullet}-1$ ($\text{density number of movers} \rightarrow  N_{\bullet\bullet}/N = 1/2$) . The dimension of this block  $N \binom{N + N_{\bullet \bullet}-2}{N_{\bullet \bullet}-1}$ scales exponentially with system size. We will show that sufficiently strong disorder 
drives an MBL transition. 

Spectral rigidity is a well known property of ergodic systems and its level spacing distribution $p(s)$ is believed to be the same as a random-matrix belonging to the same universality class~\cite{Evers2008Review,Ivan19, Luca16, Haake06}. As a result, in an ergodic phase of $\hat H_{\infty}$, $p(s)$  should be given by the Wigner surmise~\cite{Haake06, Kra09}. Instead, in a localized phase, due to an emergent weak form of integrability, $p(s)$ is given by the Poissonian distribution~\cite{Haake06, Kra09}. 
A possible way to distinguish these two cases is to study the level spacing parameter $r_n = \min \{\delta^{(n)}, \delta^{(n+1)}\}/\max \{\delta^{(n)}, \delta^{(n+1)}\}$~\cite{Oganesyan07,Atas13}, where $\delta^{(l)} = E_{l+1} - E_{l}$ are the gaps between adjacent eigenvalues $E_l$ of $\hat H_{\infty}$. In an ergodic phase the average value of $r_n$ over the energy index $n$ is given by $r_{\text{GOE}} \approx 0.5306$~\cite{Oganesyan07,Atas13} while in a localized phase $r_{\text{Poisson}} = 2\log{2}-1 \approx 0.3979$~\cite{Oganesyan07,Atas13}. 

\begin{figure}
\includegraphics[width=1.\columnwidth]{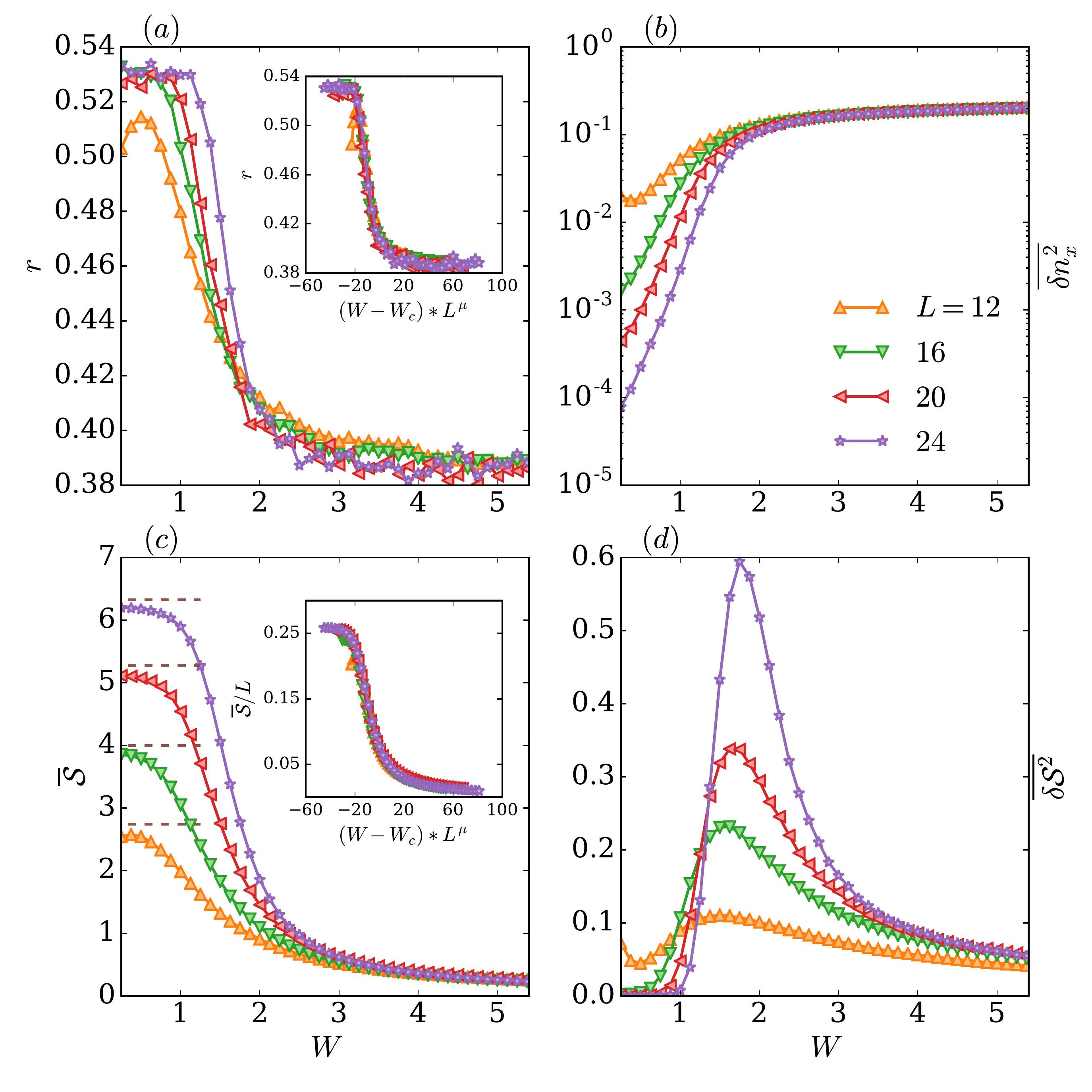}
\caption{(a): Level statistics parameter $r$ as a function of disorder strength $W$ for  several system sizes $L\in \{12,16,20,24\}$. The inset shows the finite-scaling analysis: $r$ as function of the rescaled variable $(W-W_c)L^{\mu}$ with $W_c = 2$ and $\mu=1$. (b): Fluctuations $\overline{\delta n_x^2}$ of the density operator within eigenstates in the middle of the spectrum. (c): Averaged entanglement entropy $\overline{\mathcal{S}}$ as function of $W$. The dashed line is $\overline{\mathcal{S}}\sim \log \binom{L/2}{L/4}$. Its inset shows the finite size-scaling, collapsing the rescaled $\overline{\mathcal{S}}/L$ as a function of $(W-W_c)L^{\mu}$ with $W_c = 2$ and $\mu=1$. (d): Variance of $\mathcal{S}$ within few eigenstates in the middle of the spectrum.}
\label{fig:Fig2}
\end{figure}

Figure~\ref{fig:Fig2}~(a) shows the level spacing parameter $r$ as function of disorder strength $W$ for several system sizes $L\in \{12,16,20,24\}$. We averaged $r$ over both disorder and eigenstates in the middle of the spectrum computed using shift-inverse diagonalization technique~\cite{Fra19}. At weak disorder, $r$ approaches the $r_{\text{GOE}}$ value ($W\le 1.5$) while for stronger disorder $r\approx r_{\text{Poisson}}$ ($W\ge 3$) and at intermediate values of $W$ a crossover between the two behaviors is visible.   
In order to minimize finite size effects, we use scaling techniques~\cite{Luitz15} with the aim to monitor the evolution with $L$ of the curves in Fig.~\ref{fig:Fig2}~(a). Indeed, we found the critical point $W_c$ of a putative MBL transition by collapsing the curves as a function of $(W-W_c)L^{\mu}$, see inset of Fig.~\ref{fig:Fig2}~(a). We estimate $W_c\approx 2$ and $\mu \approx 1$ for the scaling of a possible transition. 

Importantly, as one would expect, the critical point $W_c$ is smaller than the critical point for the case of finite interactions, $W_c\approx 3.5$ for $V=1$. In the limit of large interactions $V\rightarrow \infty$ the kinetic term in $\hat H_{\infty}$ is hindered  by the presence of the projectors $\{\hat P_x\}$, which forbid many of the possible hopping processes. 
We can also consider this constrained dynamics from a different point of view. Mapping the Hamiltonian $\hat H$ in Eq.~\ref{eq:full_H} to an effective Anderson problem on the Fock-space~\cite{Basko06}, in the limit $V\rightarrow \infty$ the projectors $\{\hat P_x\}$ reduce the connectivity on the Fock-space.  This effect increases the sensitivity of the system to disorder, which thus localizes more easily. This argument gives an explanation of the non-monotonic behavior of the critical point $W_c$ as function of interaction strength $V$, which has been already observed in several other works~\cite{Reich15, Scardi16}.    

We give further evidence for the existence of an MBL transition by studying the fluctuations of local observables~\cite{Beuge15, Beuge14, Alg19} 
\begin{equation}
 \delta n_x^2 = \text{Var} [ \langle E_n | \hat n_x | E_n \rangle ]_E,
\end{equation}
where the variance is taken over few eigenstates which belong to the same energy-density in the middle of the spectrum.
In an ergodic phase the expectation value of a local observable depends only on the value of the energy-density and thus $\delta n_x^2$ goes to zero in the thermodynamic limit $L\rightarrow \infty$. In fact, for states fulfilling ETH the decay is exponentially fast in $L$ $(\delta n_x^2 \sim e^{-cL}$)~\cite{Luca16}, while in a localized phase, ergodicity breaks down and eigenstates close in energy are locally different~\cite{Pal10, Luitz15}. Thus,  
we expect large fluctuations in the expectation value of the local observables $\hat n_x$, implying that $\delta n_x^2$ does not decay to zero with $L$, $\delta n_x^2 \sim \mathcal{O}(L^0)$. 
Figure~\ref{fig:Fig2}~(b) shows $\overline{\delta n_x^2}$ as function of $W$ for several system sizes. 
The fluctuations decay to zero exponentially fast in $L$ as dictated by ETH.  At large disorder $W>W_c$, no scaling with system size is visible, as one would expect in a localized phase. These results are  in agreement with our findings for the level statistic.

\begin{figure}
\includegraphics[width=1.\columnwidth]{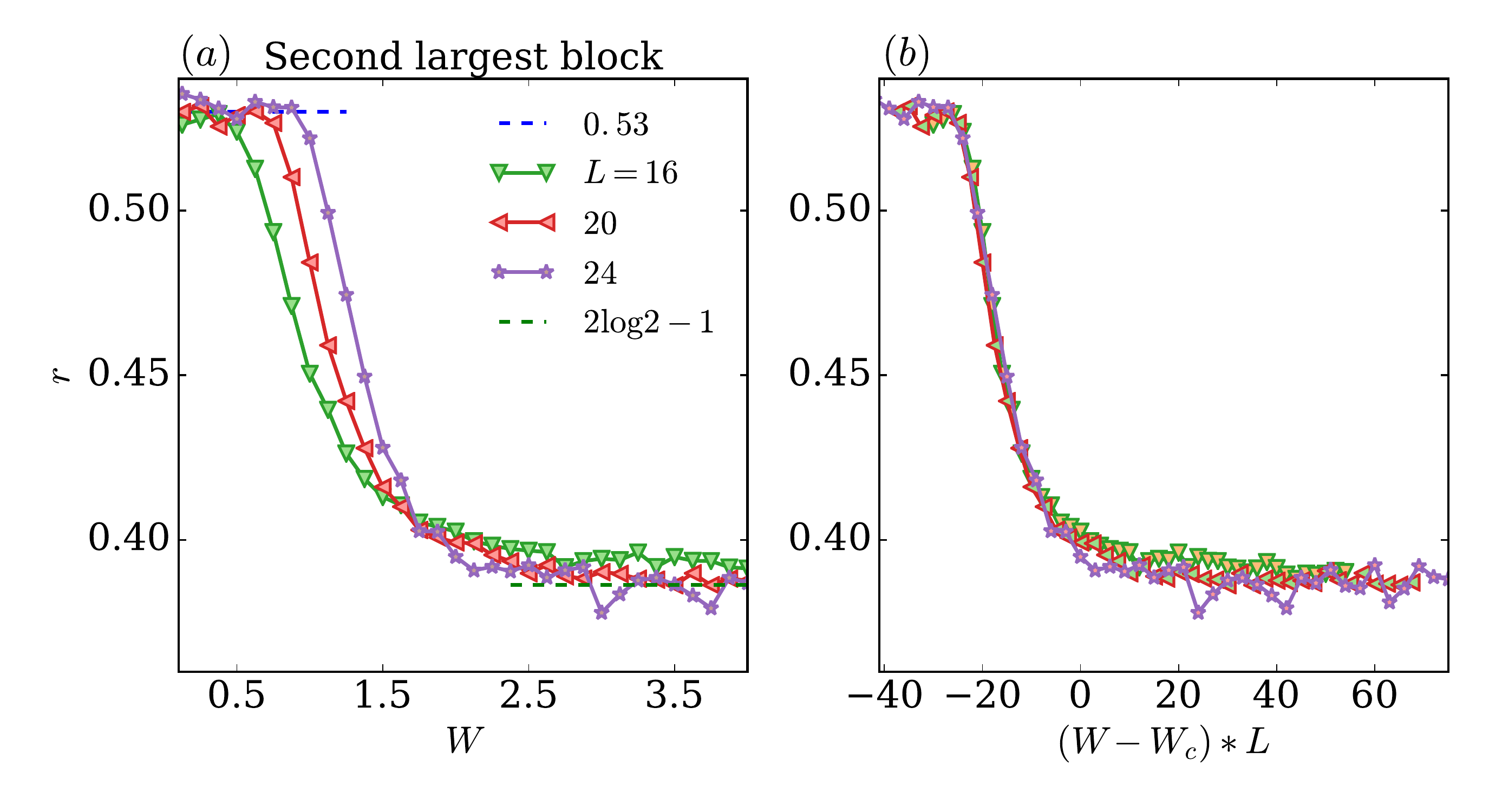}
 \caption{(a): Level statistic parameter $r$ as function of $W$ for several $L$. $r$ has been computed with eigenstates which belong to the second largest block which is characterized by $N_{\bullet\bullet}-2$ movers. The dimension of the block is given by $N \binom{N+N_{\bullet\bullet}-3}{N_{\bullet\bullet}-2}$. (b): Collapse of $r$ as function of $(W-W_c)*L^\mu$ with $W_c = 2$ and $\mu=1$.
 }
 \label{fig:Sec_MBL}
 \end{figure}

A complementary powerful method to distinguish an ergodic phase from an MBL one, is the bipartite entanglement entropy $\mathcal{S}$ of eigenstates of the Hamiltonian~\cite{Luitz15, Kjall:2014fj}. In a thermal phase the eigenstates are highly entangled and $\mathcal{S}$ follows a volume law~\cite{Luitz15, Kjall:2014fj, Pal10}, meaning that its value scales linearly with the systems size $L$. Moreover, if the system is fully ergodic at infinite temperature, the eigenstates should be described by random matrix theory implying that $\mathcal{S}$ for a typical eigenstate is given by the so called Page value $\mathcal{S}_{\text{Page}} = (L\log{2}-1)/2 + o(L)$~\cite{PageE93}. On the other hand, eigenstates in an MBL phase are only locally entangled and $\mathcal{S}$ follows an area law, $\mathcal{S} \sim \mathcal{O}(L^0)$ for one-dimensional systems~\cite{Luitz15, Kjall:2014fj, Pal10}. 

Figure~\ref{fig:Fig2}~(c) shows the entanglement entropy for eigenstates in the middle of the spectrum of $\hat H_\infty$ as a function of $W$. At weak disorder the averaged half-chain entanglement entropy  $\overline{\mathcal{S}}$, increases linearly with system size $L$, giving thus evidence that the system is delocalized. Moreover, analyzing the value of $\overline{\mathcal{S}}$ at weak disorder we find $ \overline{\mathcal{S}}\sim \log  \binom{L/2}{L/4}$. The value $\log  \binom{L/2}{L/4}{\color{blue} ^2}$ is the bipartite entanglement for a random state on the full Hilbert of $\hat H_{\infty}$~\cite{PageE93} and up to sub-leading corrections converges to the Page value $(\mathcal{S}_\text{Page}$) in the thermodynamic limit ($L\rightarrow \infty$)~\footnote{Indeed, even eigenstates that are not fully ergodic in terms of standard multifractal analysis might still reach the Page value up to sub-leading corrections. G. De Tomasi I.M  Khaymovich, V. E. Kravtsov, in preparation.}. As a consequence, at weak disorder typical eigenstates in the middle of the spectrum are ergodic. 
Instead, as the disorder is increased, $W\ge 3$, $\overline{\mathcal{S}}$ saturates with $L$, and it follows an area law $\mathcal{S} \sim \mathcal{O}(L^0)$. With the aim to understand the crossover between these two different behaviors, we collapse the curves (see inset of Fig.~\ref{fig:Fig2}~(c)) considering the rescaled entanglement entropy $\overline{\mathcal{S}}/L$ as  a function of $(W-W_c)*L^{\mu}$. In agreement with the collapse of the level spacing parameter $r$, we give an estimation for the critical point $W_c\approx 2$ with an exponent $\mu\approx 1$. 

Finally, we study the variance $\delta \mathcal{S}^2$ of $\mathcal{S}$ in eigenstates of $\hat H_\infty$~\cite{Luitz15, Kjall:2014fj}. In the vicinity of the transition, we expect large fluctuations for $\mathcal{S}$, since both thermal and localized eigenstates are considered.  Indeed, as shown in Ref.~\onlinecite{Yu16}, close to the transition the probability distribution of $\mathcal{S}$ is bimodal, with the two maxima representing ergodic, i.e. $\mathcal{S}\sim L$, and localized states, i.e. $\mathcal{S}\sim L^0$, respectively. As a consequence close to the transition we will have $\delta \mathcal{S}^2(W\approx W_c) \sim L^2$. Figure~\ref{fig:Fig2}~(d) shows $\overline{\delta \mathcal{S}^2}$ as function of $W$ for several system sizes. As expected, $\overline{\delta \mathcal{S}^2}$ close to the critical point $W_c\approx 2$,  develops a peak, which diverges with $L$, giving a further numerical evidence of the existence of an MBL transition. 

On general grounds we expect all blocks having a finite density of movers ($\# \text{movers}/N \rightarrow c\ne 0$) to have an MBL transition at some finite disorder strength $W_c$. Moreover, the critical point $W_c$ should depend on the movers density and not on the actual number of movers. As a consequence, blocks with the same density of movers should have an MBL at the same critical point. 

Figure~\ref{fig:Sec_MBL} shows the $r$ level statistic parameter as a function of $W$ computed with eigenstates which belong to one of the second largest blocks. This block is labeled by $N_{\bullet\bullet}-2$ movers and its dimension is given by $N \binom{N+N_{\bullet\bullet}-3}{N_{\bullet\bullet}-2}$. Having only one less mover with respect to the largest block, they have the same density of movers in the thermodynamic limit ($L\rightarrow \infty $). Also in this case is clear that two distinct phases exist at least for finite system sizes.  One ergodic at weak disorder and a localized one for larger $W$. The curves in Fig.~\ref{fig:Sec_MBL} can be collapsed as function of $(W-W_c)*L^\mu$ giving thus evidence of the existence of the MBL transition. In good agreement with the previous analysis we found $W_c\approx 2$ and $\mu \approx 1$ (see Fig.~\ref{fig:Sec_MBL}~(b)).  

Finally, we extended this analysis in Appendix to: (a) the presence of a quasi-periodic potential and (b) large but finite interactions for the Hamiltonian in Eq.\ref{eq:full_H}.
Furthermore, in Appendix we address to the natural question of the existence of many-body mobility edge~\cite{Luitz16}. 

\subsection{Dynamics} \label{sec:Finite_movers2}
Having demonstrated that the system shows an MBL transition, we investigate now the charge propagation focusing on the delocalized region near the MBL transition. As in the previous section we focus on the largest block with $N_{\bullet \bullet}-1$ movers. A standard description for relaxation dynamics in the system employs the density propagator~\cite{Bera17, Alg19, Reich15, Weiner19, Luitz16}  
\begin{equation}
\label{eq:density-propagator}
 \Pi(x,t) = \frac{1}{\mathcal{N}} \text{Tr}\left [ \delta \hat n _x (t) \delta \hat n_0 \right ],
\end{equation}
where $\delta \hat n_x = \hat  n_x - \frac{1}{2}$~\footnote{$\frac{1}{\mathcal{N}} \text{Tr}\left [ \hat n _x\right ]=1/2$.}, and $\mathcal{N} = N \binom{N + N_{\bullet \bullet}-2}{N_{\bullet \bullet}-1}$ is the dimension of the considered block. To monitor the dynamics of the system we define the width of $\Pi(x,t)$
\begin{equation}
 \langle X^2(t) \rangle = \sum_x x^2 \left [ \Pi(x,t) - \Pi(x,0)\right ]. 
\end{equation}
$ \langle X^2(t) \rangle $ quantifies the spreading of correlations on the system~\cite{Luitz16, Bera17,Reich15}. If the system is diffusive, $\langle X^2(t) \rangle \sim t$, while for sub-diffusive dynamics $ \langle X^2(t) \rangle \sim t^\alpha$, where $\alpha < 1$~\cite{Luitz16, Bera17,Reich15}. 
\begin{figure}
\includegraphics[width=1.\columnwidth]{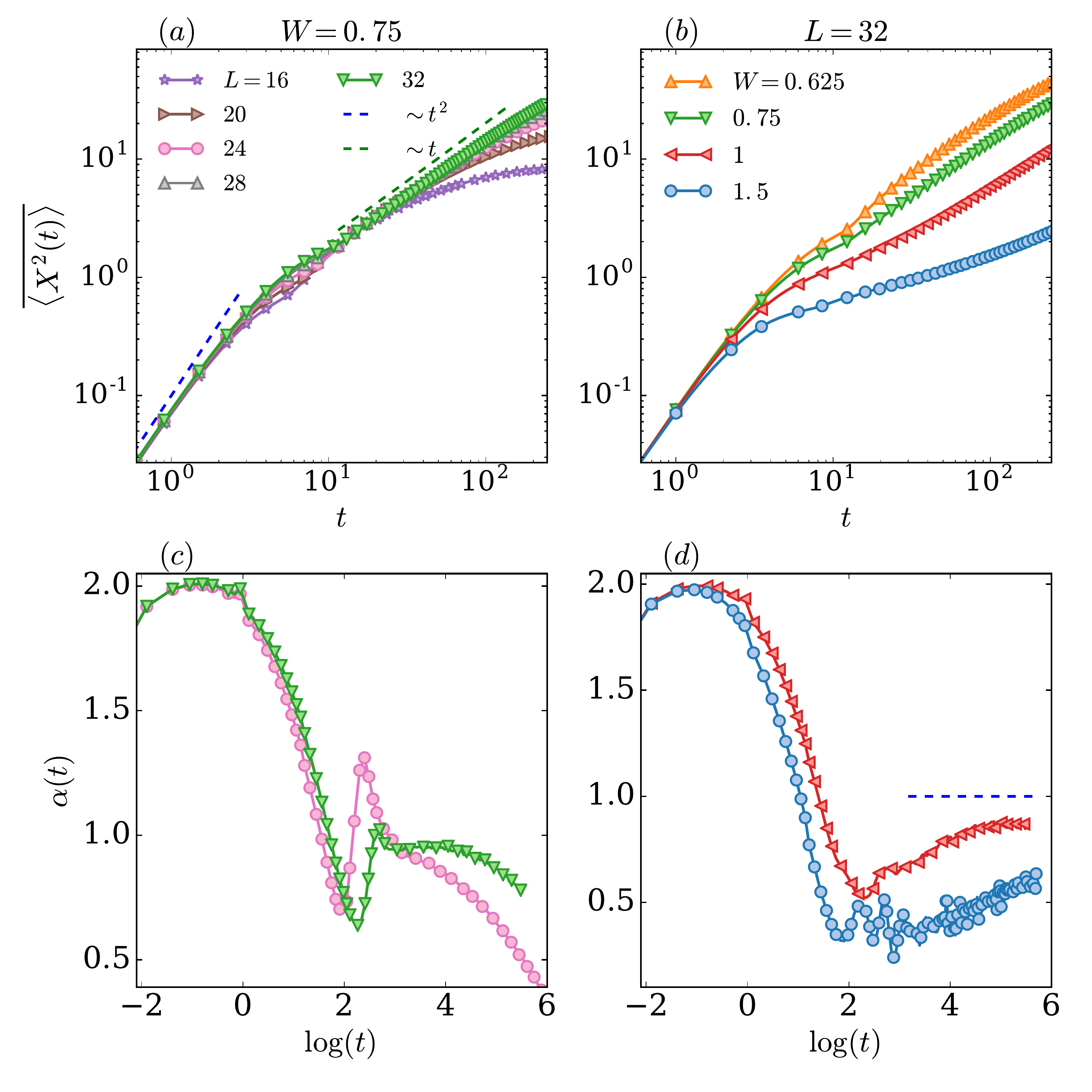}
\caption{(a): Averaged width $\overline{ \langle X^2(t)\rangle }$ of the density propagator $\Pi(x,t)$ at fixed disorder strength $W=0.75$ for several $L\in \{16, 20,24,28,32\}$. The dashed lines are guide for eyes, the blue one is the short time ballistic propagation $\sim t^2$ the red one the diffusive behavior $\sim t$. (b) Dynamical exponent $\alpha(t) = {d \log{\overline{\langle X^2(t) \rangle} }}/{d \log{t}}$ at $W=0.75$, with the enlarging plateau close to $\alpha=1$ (diffusion).    (c) $\overline{ \langle X^2(t)\rangle }$ for several $W \in \{0.625, 0.75, 1, 1.5\}$ and fixed $L=32$. (d) $\alpha(t)$ for $W=1, 1.5$ for a fixed system size $L=32$.  In these cases $\alpha(t)$ does not form a plateau at large time, instead  it increases and might approach $\alpha =1$ (dashed line) in the thermodynamic limit.}
\label{fig:Fig3}
\end{figure}

Several works have used the quantity  $\langle X^2(t) \rangle$ among others to quantify the transport in MBL systems at finite interaction strength~\cite{Luitz16, Bera17,Reich15,Weiner19, La16,Scardi16,Agar15,Pre17,Ste16,Kara09, Khait16}. These results reported the presence of sub-diffusive dynamics  on finite time scale within the ergodic phase of $\hat H$ in Eq.~\ref{eq:full_H}~\footnote{At $V=1$ a diffusive phase has been reported for $0< W/W_c<0.1$.}. However, whether this sub-diffusion is only transient or persists for asymptotically large time scale is far from being clear~\cite{Luitz16, Bera17,Ste16}. 
Moreover, the existence of rare-regions of high disorder (Griffith regions) have been invoked to explain the mechanism for sub-diffusion~\cite{Vosk15, Luitz16,AltmanTheory2015, altman2015review}. Nevertheless, the same sub-diffuse phase has been observed also for other MBL models (such as quasi-periodic~\cite{Kara09, Weiner19}, long-range hopping~\cite{Alg19}, two-dimensional systems~\cite{Bar_Lev_2016}), in which Griffith effects are suppressed, thus questioning the mechanism of this sub-diffusion propagation.   

Figure~\ref{fig:Fig3}~(a) shows $\langle X^2(t) \rangle$ as a function of time for a fixed disorder strength $W = 0.75$ and several $L\in \{16, 20, 24, 28, 32\}$. The evolution of $\Pi(x,t)$ has been computed using Chebyshev integration techniques~\cite{Bera17, Wei06}, which allow us to inspect Hilbert spaces of dimension $\approx 2,7 \cdot 10^{6}$. Also, the trace in Eq.~\ref{eq:density-propagator} has been approximated using the concept of quantum-typicality, which allows us to replace the trace with an average over random vectors~\cite{Wei06}. 

At short times settled by the hopping constant the propagation is ballistic $\langle X^2(t) \rangle\sim t^2 $. At this time scale neither disorder nor interactions have any effects and the dynamics can be approximated as a free propagation. 
As time evolves, a transient dynamics is visible in Fig.~\ref{fig:Fig3}~(a), culminating at larger times to a diffusive propagation $\langle X^2(t) \rangle \sim t$.

In order to better monitor the growth of $\langle X^2(t) \rangle $, we define the dynamical exponent~\cite{Bera17, Reich15, Luitz16}
\begin{equation}
\label{eq:dynamical_exp}
 \alpha (t) = \frac{d \log{\overline{\langle X^2(t) \rangle }}}{d \log{t}}.
\end{equation}
In diffusive systems $\alpha(t)$ develops a plateau at $\alpha=1$, while if the dynamics is sub-diffusive the plateau will be at $\alpha <1$. The study of the dynamical exponent has the advantage of identifying different time scales and sub-leading corrections that could be hidden in a fitting procedure. 
Figure~\ref{fig:Fig3}~(c) shows $\alpha(t)$ for $W=0.75$. At short times, $\alpha(t)$ reaches the value $\alpha=2$, meaning that the dynamics is ballistic, as we already discussed. At intermediate times, $\alpha(t)$ develops a plateau close to the diffusive value $\alpha =1$. 
This plateau is enlarging with increasing system size $L$, which may indicate that in the thermodynamic limit the system will be diffusive. 

Figure~\ref{fig:Fig3}~(b) shows $\langle X^2(t) \rangle $ for $L=32$ and several $W$. For smaller values of $W$ than the one just discussed, $\langle X^2(t) \rangle $ has also a diffusive behavior. 
Nevertheless, approaching the MBL transition $W_c\approx 2$, the situation is less clear. Figure~\ref{fig:Fig3}~(b) shows $\langle X^2(t) \rangle $ for $W=1$ and $W= 1.5$.
For both values of $W$, there is a time scale $t^\star$ for which 
$\langle X^2(t) \rangle $ changes curvature and might approach to the diffusive behavior $\langle X^2(t) \rangle \sim t$ at longer times. We follow the change of concavity in $\langle X^2(t) \rangle$ by studying the dynamical exponent $\alpha(t)$ for these values of disorder strength, as shown in  Fig.~\ref{fig:Fig3}~(d). 
After the ballistic propagation, i.e. $\alpha (t) \approx 2$, at short times, $\alpha(t)$ is always bounded by one, nevertheless  $\alpha(t)$ increases with time and might converge to the diffusive value $\alpha(t)=1$. 

The growth of $\langle X^2(t) \rangle $ might be characterized by two different power law behaviors 
\begin{equation}
\langle X^2(t) \rangle \sim a t^{\alpha_1} + b\frac{t}{t^\star}, \quad \alpha_1 <1.
\end{equation}
The time scale $t^\star$ defines the onset at which diffusion takes place.
Thus, at time $t\le t^\star$ the dynamics could look sub-diffusive and only for later times $t\gg t^\star$ diffusion will be completely restored. This behavior is not completely unexpected, in fact the $b/t^{\star}$ is just the diffusion constant of the system  $(D = b/t^{\star}$). Approaching the MBL transition the diffusion constant goes to zero~\cite{Basko06, gornyi2005interacting} $\lim_{W\rightarrow W_c} D(W) =0$ and the onset time scale $t^\star$ for the diffusion propagation shifts to infinity. Thus for $ 1 \le W \le 2$ we are in a regime in which the diffusive constant is extremely small. 

This regime is consistent with the theoretical prediction of Basko, Aleiner and Altshuler~\cite{Basko06} of the existence of a diffusive phase close to the MBL transition characterized by a small diffusive constant (``bad metal''). Moreover, in Ref~\onlinecite{Bera17} a similar analysis has been conducted for $\hat H$ in Eq.~\ref{eq:full_H} at finite interaction strength, which supports our findings. Therein it is shown that the dynamical exponents $\alpha(t)$ are not converged with respect to the system size and that the reported sub-diffusive propagation might be only transient.    
To summarize, we have given indication that the transport at weak disorder $W/W_c \approx 1/2$  is diffusive. At stronger disorder, i.e. closer to the MBL-transition, we see a clear crossover from a sub-diffusive dynamics to a faster propagation, compatible with diffusion in the thermodynamic limit.   

\begin{figure}
\includegraphics[width=0.9\columnwidth]{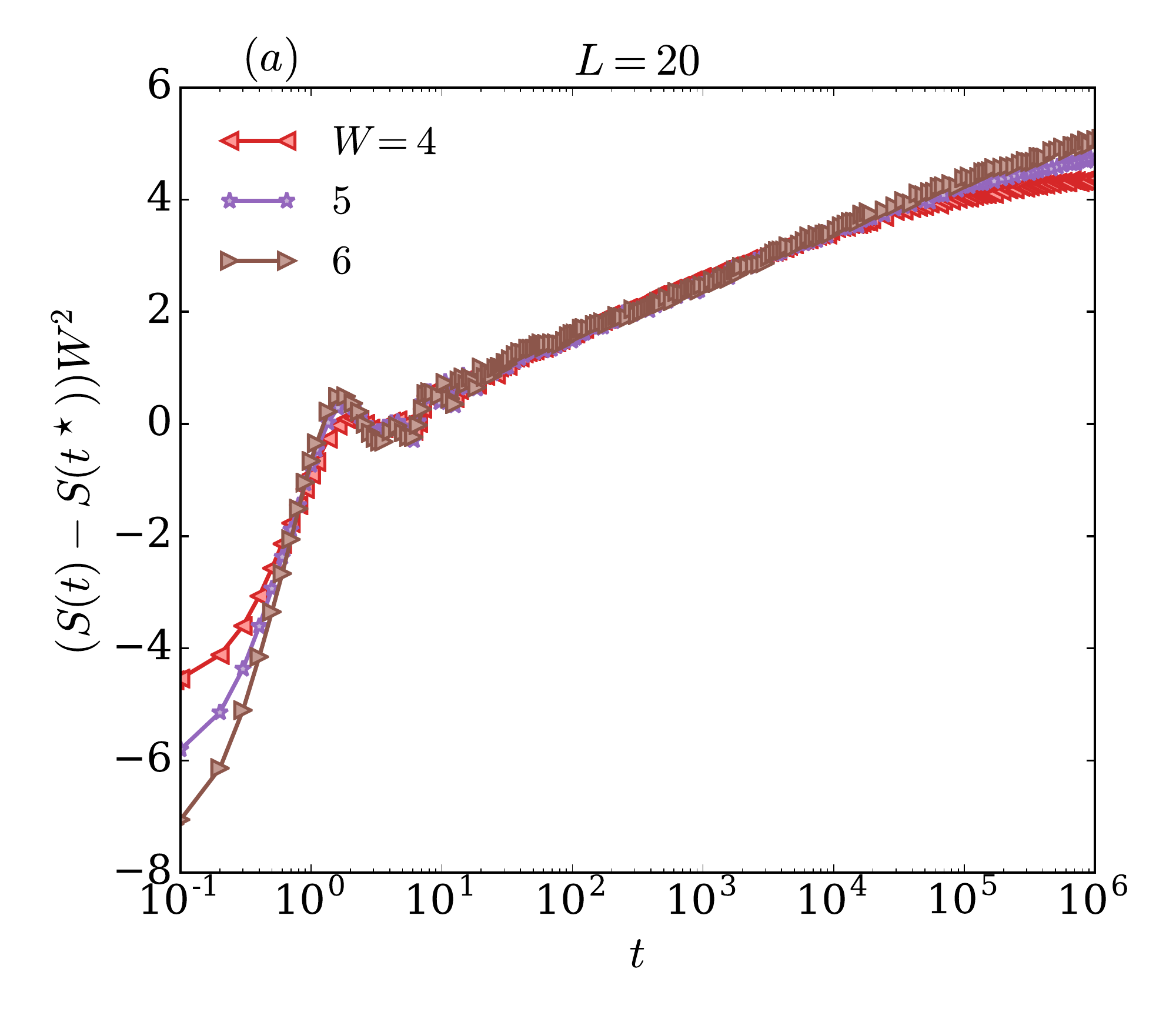}
\caption{(a): Averaged entanglement entropy after a quantum quench $\overline{\mathcal{S}}(t)$ for several $W\in\{4,5,6\}$ deep in the MBL phase and fix $L=20$. The curves have been collapsed to show $\overline{\mathcal{S}}(t) \sim \xi_{\text{loc}} \log{t}$ with $\xi_{\text{loc}} \sim 1/W^2$, where $t^\star \approx 1$.}
\label{fig:Fig4}
\end{figure}

Finally, we study the propagation of information in the MBL phase of $\hat{H}_\infty$. Although, the eigenstates in an MBL phase do not present substantial difference with the one of an Anderson insulator ($V=0$), for instance, entanglement properties, the dynamics of an MBL phase is much richer. Interactions induce a dephasing which allows slow logarithmic information propagation through the system, even though particle and energy transport is absence~\cite{GDT19, bardarson2012unbounded, serbyn2013universal}.
We compute the evolution of the bipartite entanglement entropy $\mathcal{S}(t)$ after quenching a random product state $| \psi \rangle = \prod_x \hat c_x^\dagger |0\rangle$ that belongs to the largest block of $\hat H_\infty$. Figure~\ref{fig:Fig4} shows the expected logarithmic growth of $\mathcal{S}(t)$ ($\sim \log{t}$)  for several disorder strengths $W$ deep in the MBL phase of $\hat H_\infty$, in the case in which the potential is random. The last result could be seen as a further numerical evidence of the existence of an MBL phase at strong disorder.  
Furthermore, the curves in Fig.~\ref{fig:Fig4} have been rescaled to show that the prafactor of the growth of $\mathcal{S}(t)$ could be proportional to the single particle localization length~\cite{serbyn2013universal} ($\xi_\text{loc} \sim W^{-2}$). 

\section{Conclusion} \label{sec:conclusion}
In this work we studied the $t{-}V$ disordered spinless fermionic chain in the strong coupling limit ($t/V\rightarrow 0$). 

At finite interaction strength this model is believed to exhibits an MBL transition between a thermal and a localized phase. We showed that in the limit of strong interactions strength the system is described by a kinematically constrained model with random potential. As a result, 
with increasing system size the Fock-space fragments into exponentially many disjoint blocks. Macroscopically these different blocks can be distinguished by the number of new degrees of freedom, called movers, that they can host.
We focused our investigations on two limiting cases. First, when only one mover is present in the system. Second, blocks with a finite-density of movers in the thermodynamic limit. 

In the first case we mapped the problem to an Anderson localization model with correlated disorder on the Fock-space. Using this map we showed that the system is localized for any finite amount of disorder. 
In the later case, in which the system has a finite density of movers in the thermodynamic limit, we studied numerically both eigenstates and dynamical properties. Using standard diagnostics, we provided evidence for the existence of an MBL transition at finite disorder strength.  As expected, the critical disorder strength $W_c$ is smaller compared to the case with a finite interaction strength. The reduction of the critical point is a direct consequence of the constrained dynamics, which suppresses hopping processes. 

Moreover, we studied charge relaxation in the system by employing the density propagator. We were able to access the important time scale showing that the dynamics could be diffusive on its ergodic side. In particular, close to the MBL transition we observed a transient sub-diffusive dynamics which might approach to a diffusive one at asymptotically long time. This slow transient propagation close to the transition could be the indication of the existence of the ``bad-metal'' phase, a diffusive phase characterized by a small diffusion constant~\cite{Basko06}. 

In the Appendix we extended our analysis to the case in which the disorder is generated by a quasi-periodic potential. This case is of particular interest due to recent experiments in cold-atoms~\cite{Bloch2015,schneider2015,choi2016exploring,monroe2015,Luschen17} in MBL contest. Importantly, we found the same transient sub-diffusive dynamics with a clear trend to diffusion. 

\section{Acknowledgments}
We thank  S. Bera, M. Heyl, I.M. Khaymovich and T. Rakovszky for several illuminating discussions. FP acknowledges the support of the DFG Research Unit FOR1807  through  grants  no.    PO  1370/21,  TRR80,  the Nanosystems  Initiative  Munich  (NIM)  by  the  GermanExcellence Initiative, and the European Research Council  (ERC)  under  the  European  Union’s  Horizon  2020research  and  innovation  program  (grant  agreement  no.771537).  
PS  acknowledges support from “la Caixa” Foundation(ID 100010434) fellowship grant for post-graduate studies. GDT acknowledges the hospitality of MPIPKS Dresden where part of the work was done.

\appendix 
\section*{Appendix} \label{sec:appe}

\subsection{Block structure of $\hat H_\infty$}
\begin{figure}
\includegraphics[width=1.\columnwidth]{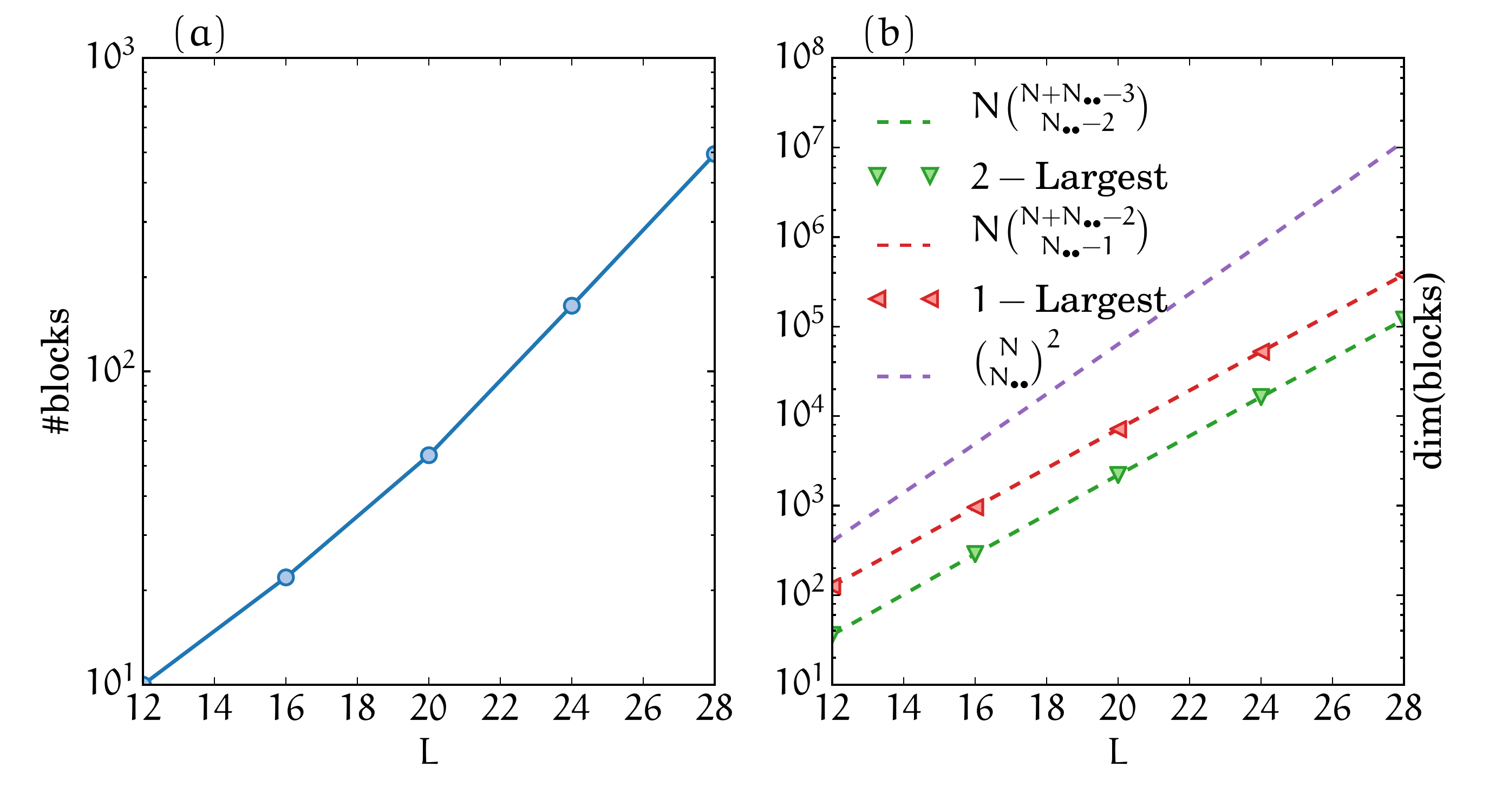}
\caption{(a): Total number of disjoint blocks in $\hat H_\infty$ as function of system size $L$ ($\# \text{blocks}\sim e^{cL}$). (b): Dimension of the largest and second largest block of $\hat H_\infty$ compared with the analytical formula $\text{dim}(\text{Largest block}) =$ $ N \binom{N + N_{\bullet\bullet}-2}{N_{\bullet \bullet}-1}$. The dashed line is the dimension of the total Hilbert space $\binom{N}{N_{\bullet\bullet}}^2$.}
\label{fig:block}
\end{figure}

In this section, we show further data on the block structure of the Hamiltonian $\hat H_\infty$ in Eq.~\ref{eq:H_inf}. In the main text, using the spin mapping introduced in Ref.~\cite{Dias2000}.  We have found that the Fock-space fragments into exponentially many blocks in $L$. We confirm this statement using numerics. Figure~\ref{fig:block}~(a) shows the total number of disjoint blocks in $\hat H_\infty$ as a function of system size $L$. The axes in Fig.~\ref{fig:block} have been chosen to underline the exponential growth $\# \text{blocks} \sim e^{c L}$. Moreover, in the main text we gave an exact formula based on the map in Ref.~\onlinecite{Dias2000} for the dimension of the block with $N_{\bullet \bullet } -1$ movers. Figure~\ref{fig:block} compared the analytical result $\text{dim}(\text{Largest block}) =$ $ N \binom{N + N_{\bullet\bullet}-2}{N_{\bullet \bullet}-1}$ with the numerical one, finding a perfect match. 

In the remainder of this section we show that the total number of blocks \textit{within} the subspace of $N=L/2$ and $N_{\bullet\bullet}=N/2$ scales exponentially in system size, without using the spin mapping~\cite{Dias2000}. To this end we find lower bounds for the number of disconnected blocks for more than one mover. Doing so, we estimate the number of blocks to scale faster than $\frac{8}{L}3^{L/8}$.

\begin{figure}
\centering
\includegraphics[width=1\linewidth]{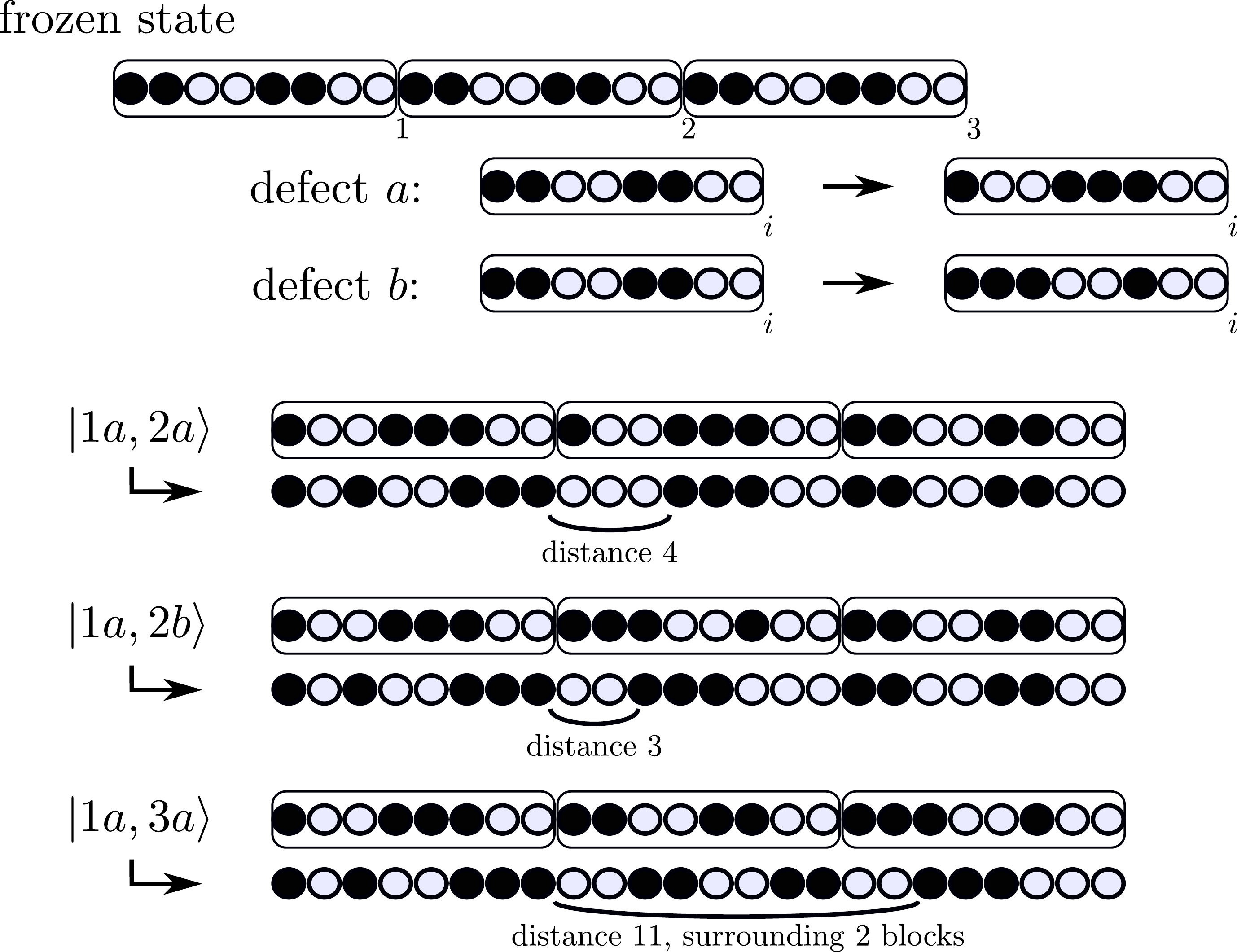}
\caption{Illustration of the analytical argumentation of having  an exponential fragmentation of the Fock-space. Two different defects ($a$ and $b$) are performed to domains of the frozen state, which results in disconnected states, see main text.}
\label{fig:exp_scaling}
\end{figure}

In order to obtain sector with more than just one mover, we start from one of the frozen states, see Fig.~\ref{fig:exp_scaling}. We then divide the whole chain in $L/8$ regions containing eight adjacent sites each, such that each region  $i\in {1,...,L/8}$ contains two blocks of two particles separated by two empty sites. We define two kind of perturbations (defects) $a$ and $b$, which act only on an individual domain $i$. Concretely, $a$ attaches the right particle of the left block to the left side of the right block and $b$ attaches the left particle of the right block to the right side of the left block, see Fig.~\ref{fig:exp_scaling}. Note that both perturbations create a block of three particles and an unbound separate particle. 

For each of the $L/8$ regions, we can then perform three different actions: defect $a$, $b$, and the identity (no action) (see Fig.~\ref{fig:exp_scaling}), yielding $3^{L/8}$ different states. Now it remains to show which of those states may be connected with each other by the action of $\hat H_{\infty}$. To this end, let us denote these states by $\ket{i\alpha, j\beta, \ldots}$, where perturbation $\alpha$ has been performed on region $i$, etc. For instance, $\ket{1a,2a}, \ket{1a,2b},$ and $\ket{1a,3a}$ each experienced two defects and are shown in Fig.~\ref{fig:exp_scaling}.
As each defect creates exactly one separate particle and further separate particles cannot be created during the dynamics, see the rules shown in Fig.~\ref{fig:rules}), a minimum criterion for a possible connection between two of our constructed states via the Hamiltonian $\hat H_{\infty}$ is the equality of the number of performed defects.

Now, let us move or tunnel all separate particles (movers) to the left end of the chain, while, according to the hopping rules, domains of multiple particles move two sites to the right for each crossing mover. The resulting states are equal to each other on the left side of the chain, where the movers are, but the distribution of the blocks on the right side of the chain differs, see Fig.~\ref{fig:exp_scaling}. Specifically, the distances between the blocks of three particles, which are side products of the defects, depends on \emph{where} and \emph{which} defect has been performed. As it is impossible for an individual block to move or even interchange its position with a different block, all such states are disconnected from each other unless they merge by cyclic permutations. With $L/8$ possible cyclic permutations of a set of defects, we end up with $\frac{8}{L}3^{L/8}$ disconnected sectors that emerge from the above defined defects on the frozen state. This gives a lower bound for the number of disjointed blocks. Note that this simple construction does not even take into account that blocks of sizes larger than three sites are possible, which quickly increases the number of disconnected sectors in the Hilbert space.

\begin{figure}
 \includegraphics[width=1.\columnwidth]{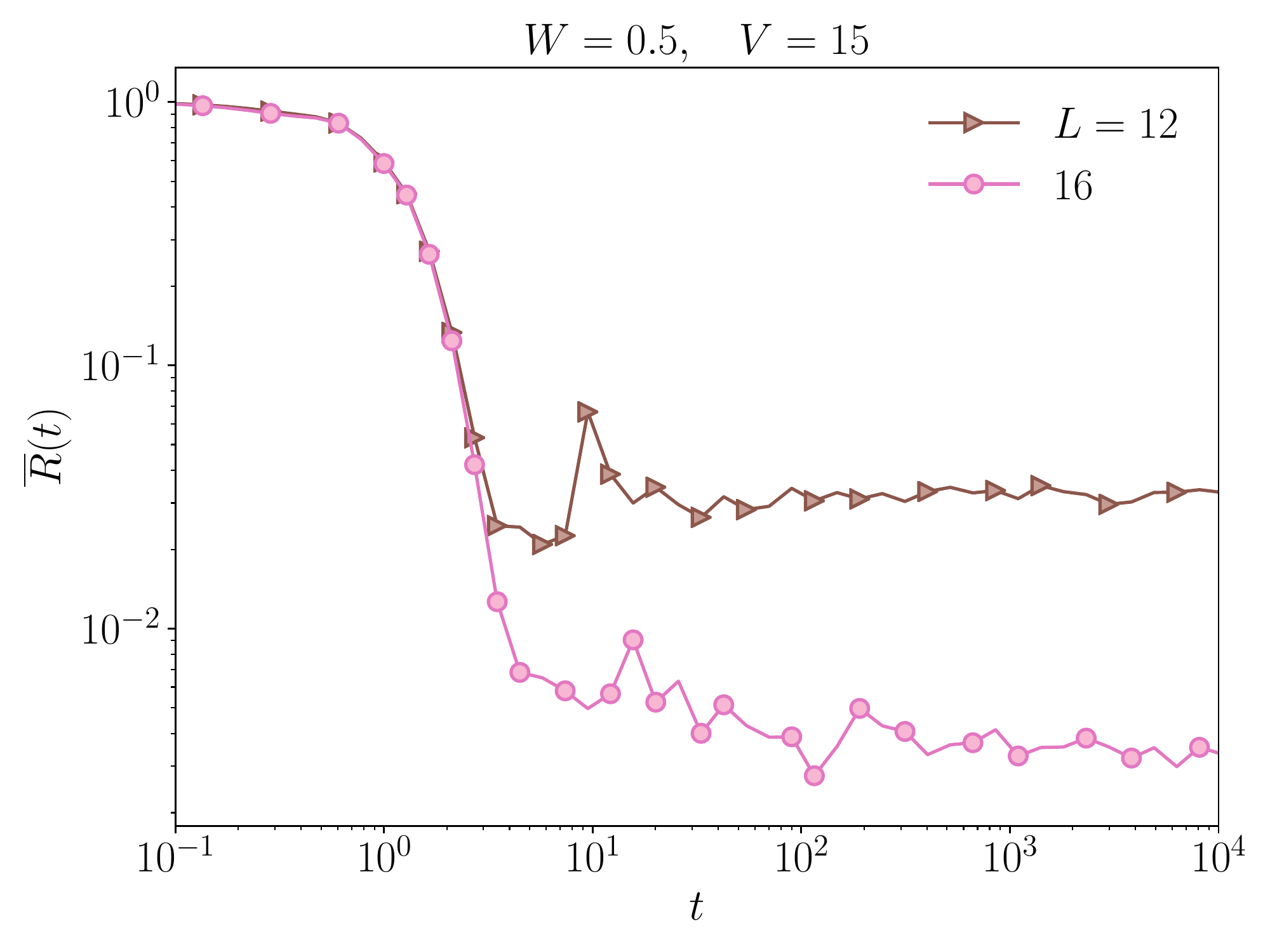}
 \caption{ $\overline{R}(t) = \overline{|\langle j | e^{-i\hat H t} |j \rangle |^2}$ at large interaction strength $V=15$ and weak disorder $W=0.5$ and $L\in \{	12,16\}$. 
 $\overline{R}(t)$ has been averaged over disorder. The initial state $|j\rangle$ have been taken from a block of $\hat H_\infty$ with a $N_{\bullet \bullet}-1$ movers.}
 \label{fig:R_finite_density}
 \end{figure}
 \subsection{$t{-}V$ model with random potential}
 In this section we provide further data for the Hamiltonian $\hat H$ in Eq.~\ref{eq:full_H} at large $V\gg t,W$ and for the limit of strong coupling $t/V \rightarrow 0$ in Eq.~\ref{eq:H_inf}. 
 
 In the main text we have shown that in the limit of strong coupling $t/V\rightarrow 0$  the return probability $R(t)$ in Eq.~\ref{eq:return_prob} does not 
 goes to zero if the system is initialized with a Fock state that host only one mover. 
 
 Moreover, we have given evidence that the limit $t/V\rightarrow 0$ could be used 
 to shed light on the finite time dynamics of $\hat H$ with $V$ large but finite. Indeed, in the main text we have presented converged data with time, showing that $R(t)$ computed with 
 $\hat H$ with large $V$ has a slow dynamics and the system is almost-localized up to time scales of order $\sim V^2$. Nevertheless, our theory predicts that this pre-thermal behavior should be 
 absent once $R(t)$ is computed starting with a state with a finite density of movers (at least at weak disorder). 
 
 Figure~\ref{fig:R_finite_density} shows $\overline{R}(t) = \overline{|\langle j | e^{-i\hat H t} |j \rangle |^2}$ at large interaction strength and weak disorder but 
 starting with an initial state with a finite density of movers ($\# \text{movers} = N_{\bullet \bullet} -1$). 
 As expected, in this case $\overline{R}(t)$ decays to zero faster than the case considered in the main text. 
 Importantly, it does not develop any pre-thermal plateau. This result is  consistent  with the analysis that we provided in Sec.~\ref{sec:Finite_movers}.
 
 \begin{figure}
 \includegraphics[width=1.\columnwidth]{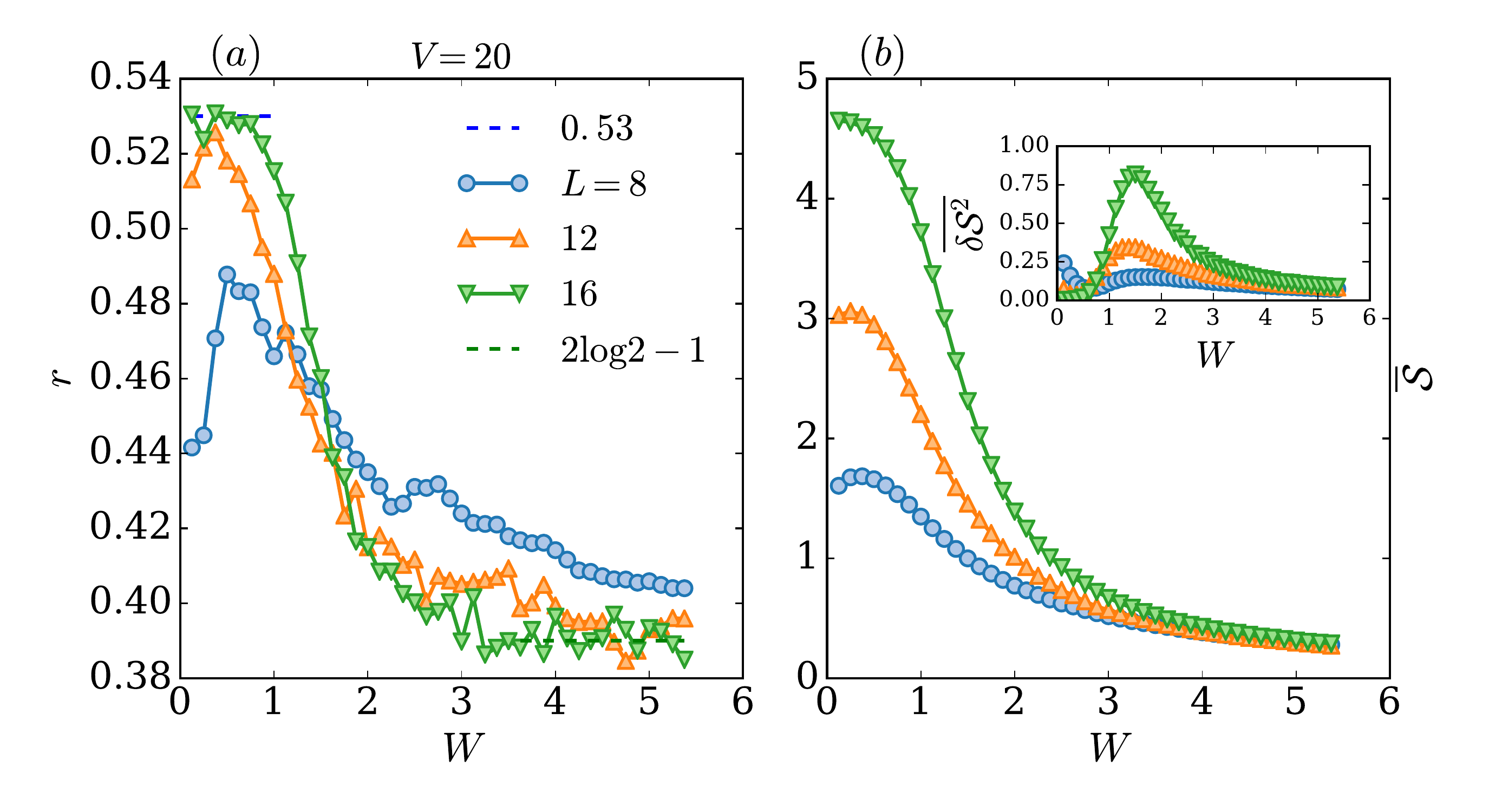}
 \caption{(a): Level statistics parameter $r$ as a function of disorder strength $W$ for eigenstates in the middle of the spectrum of $\hat H$ in Eq.~\ref{eq:full_H} with $V=20$. (b):
 Averaged bipartite entanglement entropy $\overline{\mathcal{S}}$. Its inset shows the variance $\overline{\delta \mathcal{S}^2}$ of $\mathcal{S}$ within eigenstates in the middle of the spectrum. 
 }
 \label{fig:V_20}
 \end{figure}
 
 Now we address the question of the ergodic properties of $\hat H$ for $V$ large but finite. Figure.~\ref{fig:V_20} shows the level statistic parameter $r$ and the averaged bipartite entanglement entropy $\overline{\mathcal{S}}$ for eigenstates in the middle of the spectrum of $\hat H$ in Eq.~\ref{eq:full_H} with $V=20$. The eigenstates of $\hat H$ have been computed using shift-inverse diagonalization techniques. At finite strength $V$ the block structure that we have discussed in the main text is lost. As a result, the reachable system size are smaller than the strong coupling limit case. 
 
 Both quantities in Fig.~\ref{fig:V_20} show the typical behavior as a function of $W$ for a system having an MBL transition~\cite{Pal10, Luitz15}. At weak disorder the system thermalizes, while at larger disorder it has the salient properties of a localized phase, i.e. Poisson level statistics ($r\approx 0.39$) and area law entanglement 
 ($\mathcal{S} \sim \mathcal{O}(L^0)$).
 
 Moreover, the inset in Fig~\ref{fig:V_20}~(b) shows the variance $\overline{\delta \mathcal{S}^2}$ of $\mathcal{S}$ within eigenstates in the middle of the spectrum, with the typical diverges around the critical point $ \delta \mathcal{S}^2(W\approx W_c)\sim L^2$.
 Due to the limitation of system size, we were not able to estimate reliability the critical point $W_c(V=20)$ of the MBL transition. However, on a qualitative basis and for these system sizes ($L\le 16$) the crossover between the two phases seems consistent with the limit that we studied in the main text ($t/V\rightarrow 0$). 
  \begin{figure}
 \includegraphics[width=1.\columnwidth]{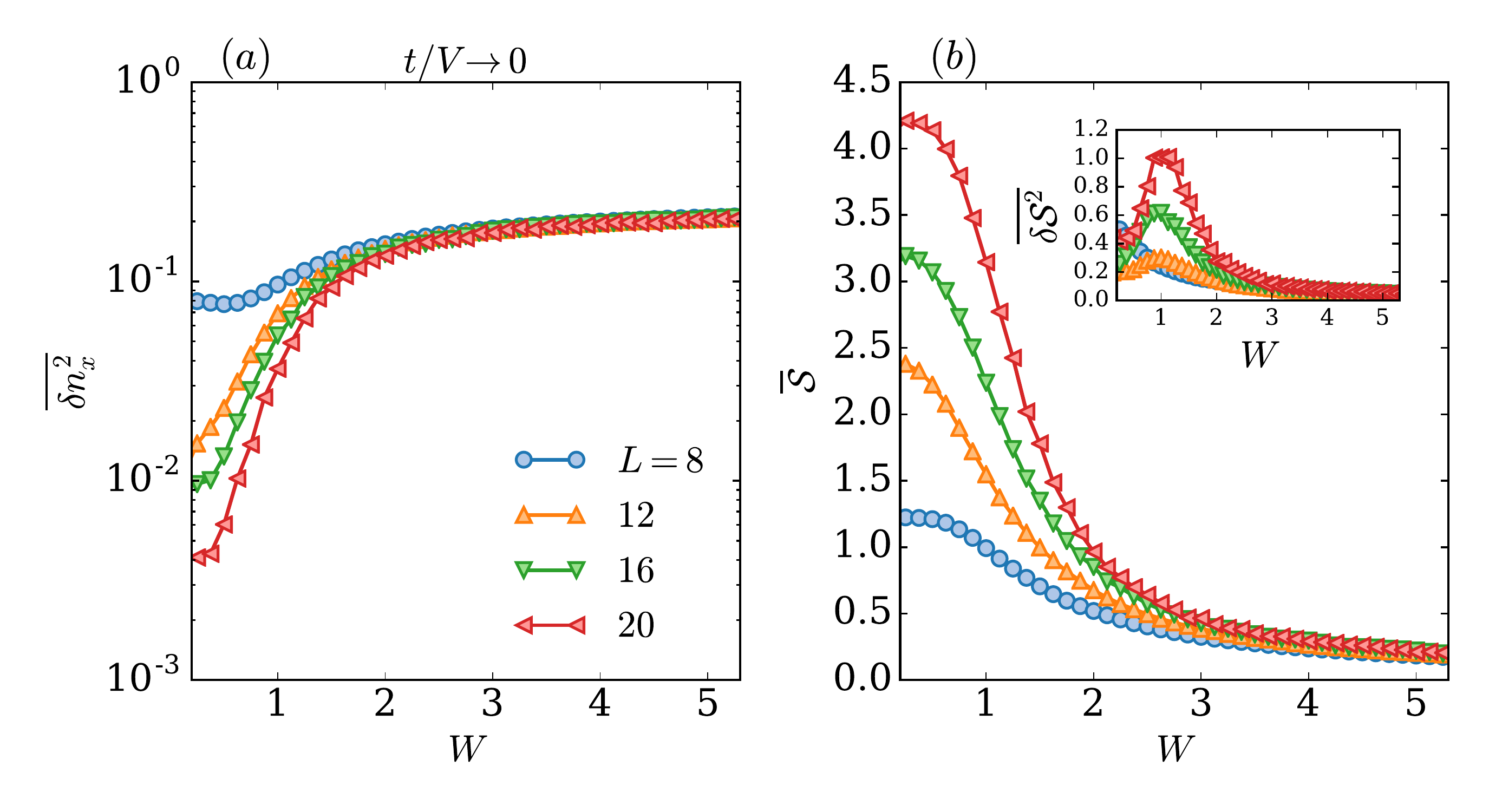}
 \caption{(a): Fluctuations $\overline{\delta n_x^2}$ of a local observable for eigenstates of $\hat H_\infty$ in Eq.~\ref{eq:H_inf} without restricting it to a specific block. (b):
 Averaged bipartite entanglement entropy $\overline{\mathcal{S}}$. Its inset shows the variance $\overline{\delta \mathcal{S}^2}$ of $\mathcal{S}$ within eigenstates in the middle of the spectrum. 
 }
 \label{fig:V_inf_W}
 \end{figure}

   \begin{figure}
\includegraphics[width=1.\columnwidth]{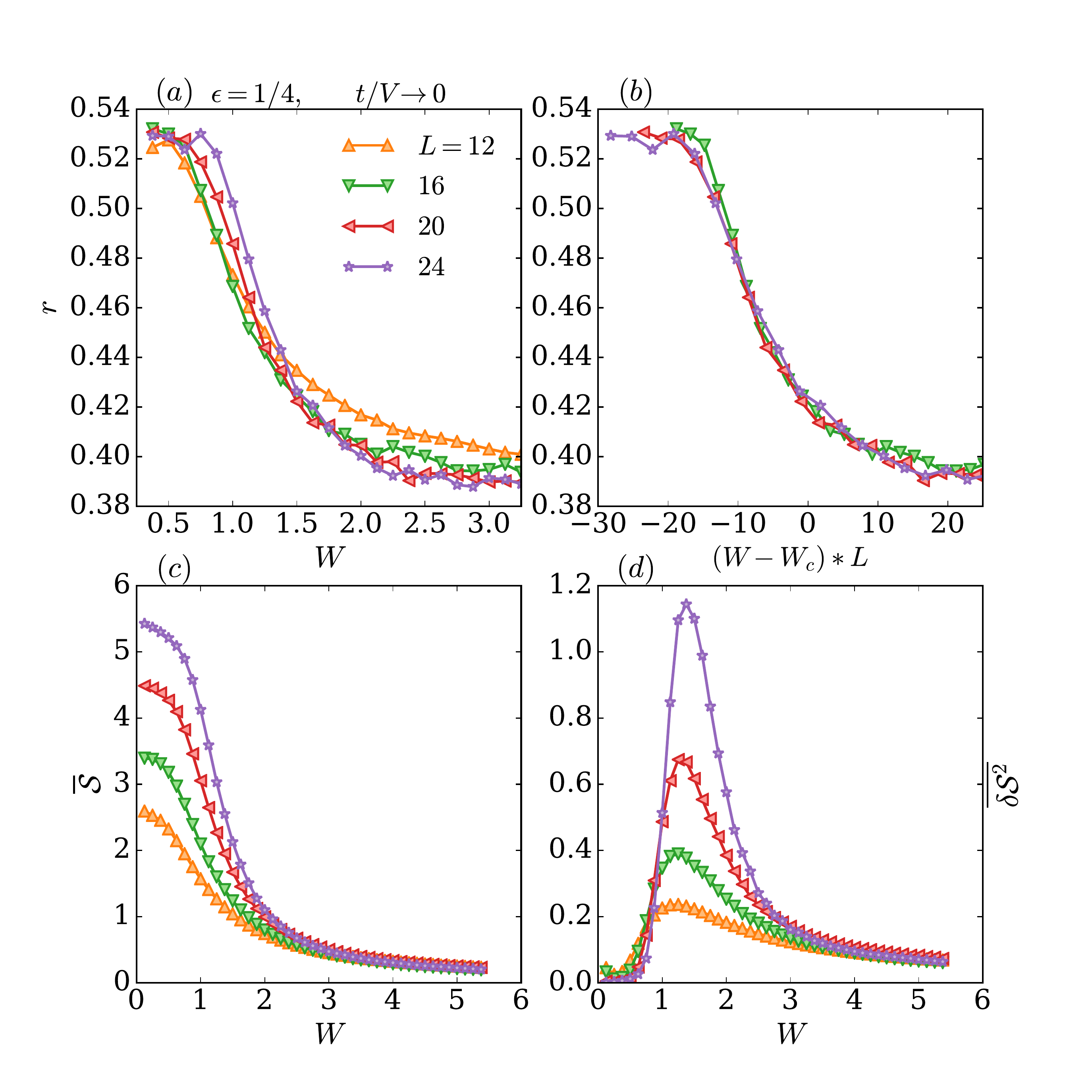}
 \caption{(a): $r$ level spacing parameter as function of disorder strength $W$.  (b): Collapse of $r$ as a function of $(W-W_c)*L^{\mu}$ with $W_c = 1.55$ and $\mu =1$. (c): Averaged bipartite entanglement entropy $\overline{\mathcal{S}}$. (d): Variance $\overline{\delta \mathcal{S}^2}$ of $\mathcal{S}$ within eigenstates of $\hat H_\infty$. All the panels have been computed using eigenenergies or eigenstates at energy density $\epsilon = 1/4$. 
 }
 \label{fig:mobility_V_inf}
 \end{figure}

 In the main text we have studied the limit $t/V\rightarrow 0$ of $\hat H$ in Eq.~\ref{eq:full_H} and we focus on two particular blocks. Here, we consider the eigenstates properties of $\hat H_\infty$ in Eq.~\ref{eq:H_inf} but 
 without restricting it to a specific block. This analysis allows us to understand what is the overall behavior of the system on disorder strength $W$. We expect that blocks with a finite density of movers will show an MBL transition, which depends on the movers density $\sim \# \text{movers} / L$. On the other hand blocks with a zero density of movers in the limit $L\rightarrow \infty $ should be  localized for any amount of $W$. Thus, it is natural to pose the question of the overall behavior. 
 
 Due to the block structure of $\hat H_\infty$ it is clear that its level statistic is Poissonian ($r\approx 0.39$), but this is just an artefact of the considered limit. Nevertheless, the eigenstates of $\hat H_\infty$ have a delocalization-localization transition. Figure~\ref{fig:V_inf_W} shows the fluctuation of a local observable $\overline{\delta n_x^2}$ and the entanglement entropy $\overline{\mathcal{S}}$. $\overline{\delta n_x^2}$ and $\overline{\mathcal{S}}$ have been computed using eigenstates of $\hat H_\infty$ in the middle of the spectrum. 
 Both quantities give indication of the existence of two distinct phases at least for finite systems. At weak disorder  $\overline{\delta n_x^2}$ decays to zero exponentially fast with $L$ and $\overline{\mathcal{S}}$ shows a volume law, while at larger disorder both $\overline{\delta n_x^2}$ and $\overline{\mathcal{S}}$ saturates with $L$. It is important to notice that the value of the entanglement entropy is below the value predicted by random matrix $\mathcal{S}_{\text{Page}} = (L\log(2)-1)/2$, although it has a volume law ($\mathcal{S} \sim L$). Nevertheless, we cannot rule out that this difference is only  to due sub-leading terms. 
 
 \subsection{Many-body mobility edge in the strong limit case $t/V\rightarrow 0$}
 
The Hamiltonian $\hat H$ in Eq.~\ref{eq:full_H} at finite interaction strength $V$ is believed to have a many-body mobility edge (MBME)~\cite{Luitz16}, meaning that on an extensive region of the phase diagram ($W<W_c$),  the energy spectrum separates between localized eigenstates at low energy density and ergodic ones in the middle of the spectrum. At larger value of disorder strength $W>W_c$ the system is in a fully MBL phase and all its eigenstates are localized. 

It is important to warn the reader that both the position of the  MBME and its existence are still under debate~\cite{DeRock16, Bera17, Lev19}. 
In general, finding the boundaries of the MBME on the phase diagram is an extremely hard task, since it involves a detailed analysis of the model at several energy scales. 

In this section, we address the natural question of the existence of the MBME in the limit of strong interactions $t/V\rightarrow 0$. Although, our main goal is not to describe the entire phase diagram, but only to provide indication that the critical point of the MBL transition might depend on the energy density.  In particular, we will study both spectral and eigenstates properties of $\hat H_\infty$ in Eq.~\ref{eq:H_inf}  but at a different energy density than the one we considered in the main text. 

Figure~\ref{fig:mobility_V_inf}~(a) -- (d) shows several standard MBL diagnostics (i.e. $r$, $\mathcal{S}$) for eigenenergies and eigenstates of $\hat H_\infty$ restricted to a block with $N_{\bullet\bullet}-1$ movers at energy density $\epsilon \equiv (E-E_{\text{min}})/(E_{\text{max}}-E_{\text{min}}) = 1/4$. In the main text we studied ergodic properties of $\hat H_{\infty}$ in the middle of the spectrum $\epsilon = 1/2$, finding an MBL transition at $W_c (\epsilon = 1/2) \approx 2$. 

Figure~\ref{fig:mobility_V_inf} should be compared with Fig.~\ref{fig:Fig1} in the main text. As expected, weaker disorder is needed to localize the system (i.e. $r\approx r_{\text{Poisson}}$ and $\overline{\mathcal{S}}\sim \mathcal{O}(L^0)$) than the case considered in the main text ($\epsilon=1/2$). As we did in the main text,  we collapse the curves of the $r$ level spacing parameter for several systems sizes $L$  (see Fig.~\ref{fig:mobility_V_inf}). We estimate the critical point of a putative MBL transition at $W_c(\epsilon=1/4) \approx 1.55 < W_c(\epsilon = 1/2)$.

 \subsection{$t{-}V$ model with quasi-periodic potential}
 In this section we study the Hamiltonian $\hat H$ in Eq.~\ref{eq:full_H} in the case in which the disorder is generated by the presence of a quasi-periodic potential. 
 $\mu_x = \cos(2\pi \sigma x + \alpha)$, where $\sigma = (1+\sqrt{5})/2$ is the Golden ration and $\alpha$ a random phase uniformly distributed between $[0, 2\pi]$. 
 Its non-interacting limit ($V=0$) is known as  Aubry-Andr\'e (AA) model~\cite{Aubry80}. 
 The AA model has a metal-insulator transition between extended to localized wavefunctions at $W_c=1$~\cite{Aubry80, Giu17, Var17, Iyer13}. 
 Numerical and experimental works have shown that the 
 AA-model at finite interaction has also an MBL transition~\cite{Iyer13,Sirker19,Weiner19,Karr17, Bloch2015,Zhang18,Vedika17,Luschen17}.  
 
 As in the main text we consider the strong coupling limit $t/V\rightarrow 0$, obtaining $\hat H_{\infty}$ in Eq.~\ref{eq:H_inf} but with $\mu_x = \cos(2\pi \sigma x + \alpha)$. 
 Moreover, we focus on the two limiting cases: first, the case in which the system hosts only one mover. 
 Secondly, the system with $N_{\bullet \bullet} -1 = L/4 -1$ movers. 
 
\begin{figure}
 \includegraphics[width=1.\columnwidth]{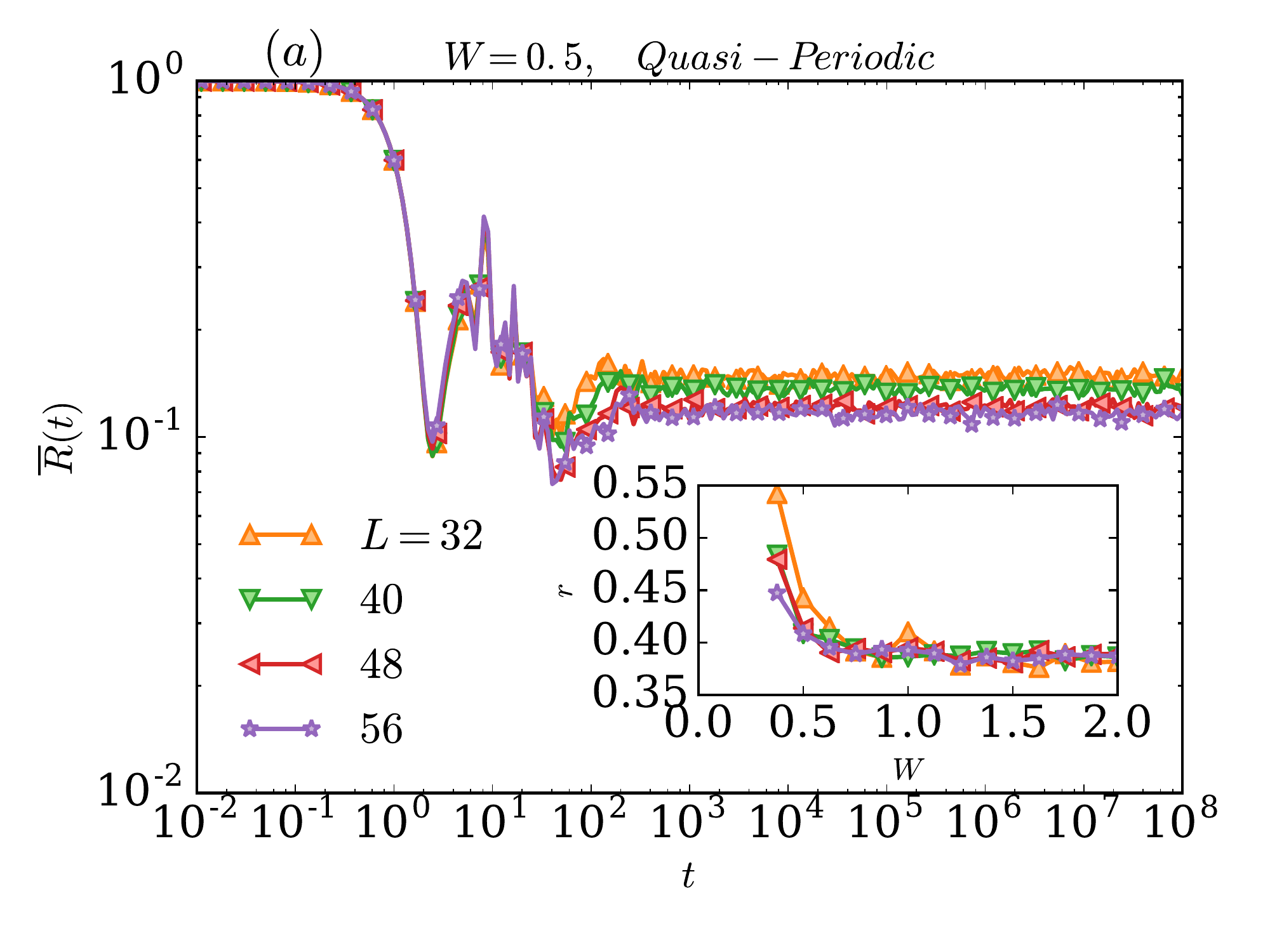}
 \caption{(a): $\overline{R}(t)$ for $W=0.5$ (strength of the quasi-periodic potential) and several system sizes $L\in \{32, 40, 48, 56\}$. As for the case with random potential (see main text) $\overline{R}(t)$ does not decay to zero, meaning that the system is localized. Its inset shows the level statistics parameter $r$ defined in the main text. $r$ approaches the Poissonian value (dashed line) increasing system size. }
 \label{fig:R_QP}
 \end{figure}
We start our discussion from the case in which $\hat H_\infty$ is restricted to the blocks with one mover.  As we already discussed its Fock-space is composed by $N^2= (L/2)^2$ states.
Figure~\ref{fig:R_QP} shows the averaged return probability $\overline{R}(t)$ 
defined in the main text in Eq.~\ref{eq:return_prob} for a fixed $W=0.5$ and several system sizes $L\in \{32,40,48,56\}$. 
As in the case with a random potential, $\overline{R}(t)$ does not relax to zero with time but it saturates to a finite positive value, 
giving thus indication that the system is localized. 
Indeed, in this block the model can be mapped to a single-particle hopping problem subjects to the following potential 
$\chi_j = W \sum_x \cos(2\pi \sigma x + \alpha) \langle j | \hat n_x | j \rangle$. 
In the limit of large system size $L$, $\chi_j$ is a sum of an extensive number of dephased cosine functions and thus to first approximation $\{\chi_j\}$ can be considered as random variables.
As a result, we expect to have localization for any value of $W$ ($W\ne 0$). 
It is interesting to note that the single-particle transition in the AA-model ($W_c(V=0)=1$) is washed out since the system is localized for any $W$ (in Fig.~\ref{fig:R_QP} $W=0.5 < W_c(V=0)$). 
The inset of Fig~\ref{fig:R_QP} shows the level statistic parameter $r$. As one would expect, being the system localized, $r$ approaches the Poisson value 
$r_{\text{Poisson}} = 2\log{2}-1$ (dashed line) for any $W$. 
\begin{figure}
 \includegraphics[width=1.\columnwidth]{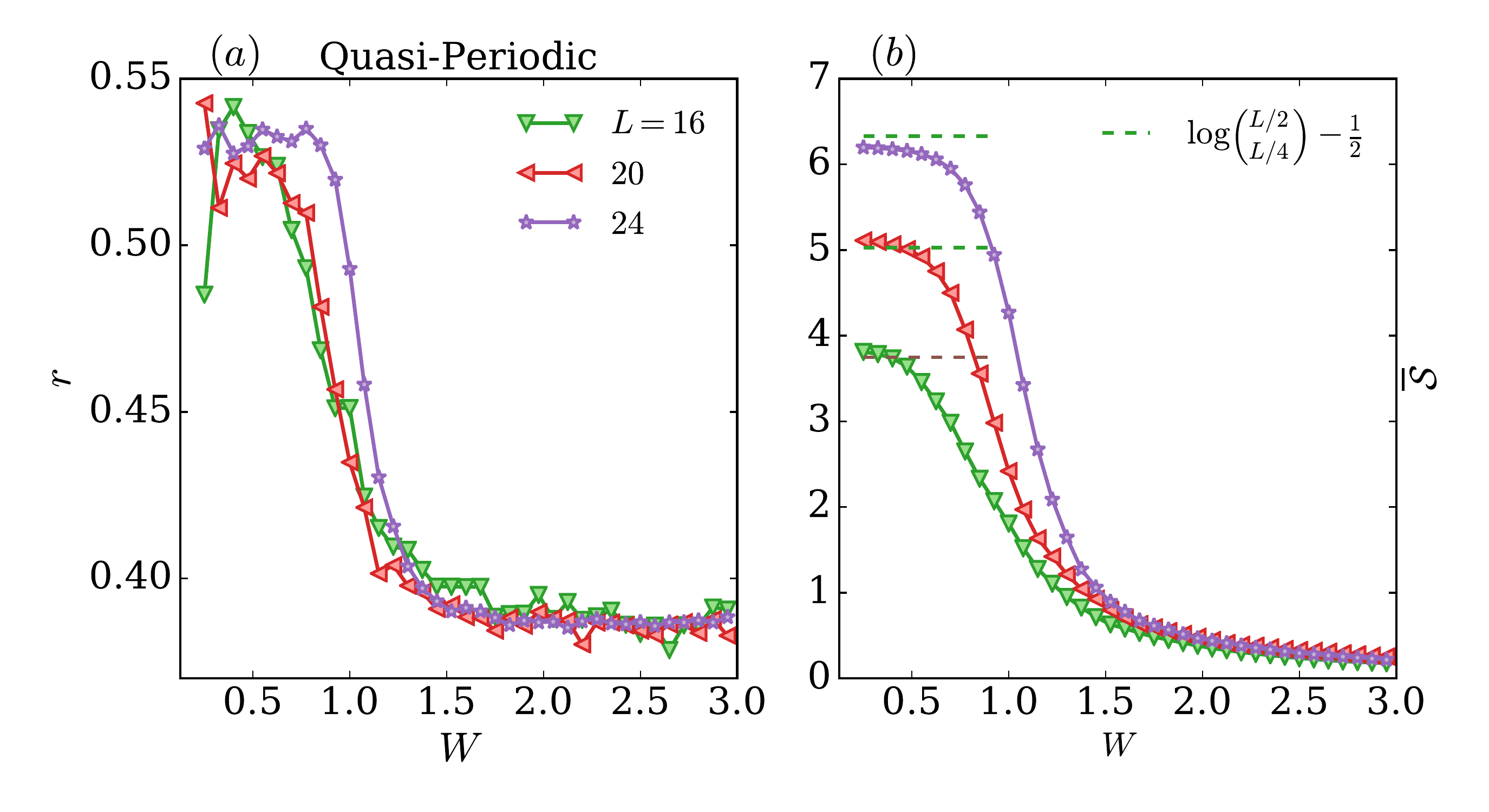}
 \caption{(a): Level statistics parameter $r$ as a function of the strength $W$ for  several system sizes $L\in \{16,20,24\}$. 
 At small $W$, $r$ converges to $r_{\text{GOE}}\approx 0.53$, while at larger $W$ ($W>2$), $r$ takes the Poisson value. 
 (b): Averaged entanglement entropy $\overline{\mathcal{S}}$ as function of $W$. The dashed line is $\overline{\mathcal{S}}\sim \log \binom{L/2}{L/4}$. }
 \label{fig:v_inf_static_QP}
 \end{figure}
 
Now we turn to the case of $N_{\bullet \bullet} -1 = L/4 -1$ movers. 
In this case the dimension of the Hilbert space is given by  $N \binom{N+N_{\bullet\bullet} -2 }{N_{\bullet \bullet} -1}$ 
and it grows exponentially fast in $L$. 
First, we focus on both spectral and eigenstates properties showing that the system might have an MBL transition. 
Second, we investigate the dynamics of $\hat H_{\infty}$ with quasi-periodic potential within its ergodic phase.

Figure~\ref{fig:v_inf_static_QP}~(a) shows the level spacing parameter $r$ as function of $W$ for several system sizes $L\in \{16,20,24\}$. 
We averaged $r$ over both the random phase $\alpha$ in $\{\mu_x\}$ and eigenstates in the middle of the spectrum. 
At small $W$, $r$ approaches the $r_{\text{GOE}}$ value while for larger $W$, $r\approx r_{\text{Poisson}}$ and at intermediate strengths a crossover between the two behaviors is visible. 

Figure~\ref{fig:v_inf_static_QP}~(b) shows the averaged bipartite entanglement entropy $\overline{\mathcal{S}}$ for eigenstates of $\hat H_\infty$ in the middle of the spectrum  as a function of $W$.
At weak disorder $\overline{\mathcal{S}}$ has a volume-law, meaning that it increases linearly with system size $L$. 

However, at larger $W$ ($W\ge 2$) $\overline{\mathcal{S}}$ saturates with $L$, and it follows an area law ($\mathcal{S} \sim \mathcal{O}(L^0)$).
As we did for the case with random potential (main text), we analyze the value of $\overline{\mathcal{S}}$ at small $W$, we find that  also in this case 
the scaling $ \overline{\mathcal{S}}\sim \log  \binom{L/2}{L/4}$, which converges to the Page value up to sub-leading corrections.

Summarizing, Fig.~\ref{fig:v_inf_static_QP} give indication of the existence of two distinct phases, 
 ergodic and localized. Using the same scaling analysis techniques that we used in the main text, 
we estimate the critical point for the quasi-periodic case, $W_c \approx 1.6$. 
As one would expect the found critical point ( $W_c \approx 1.6$) is larger than the critical point of the non-interacting AA model ($W_c(V=0)$) 
but smaller than  the one at finite interaction strength ($W_c(V=1) \approx 3.5$). 

\begin{figure}
\includegraphics[width=1.\columnwidth]{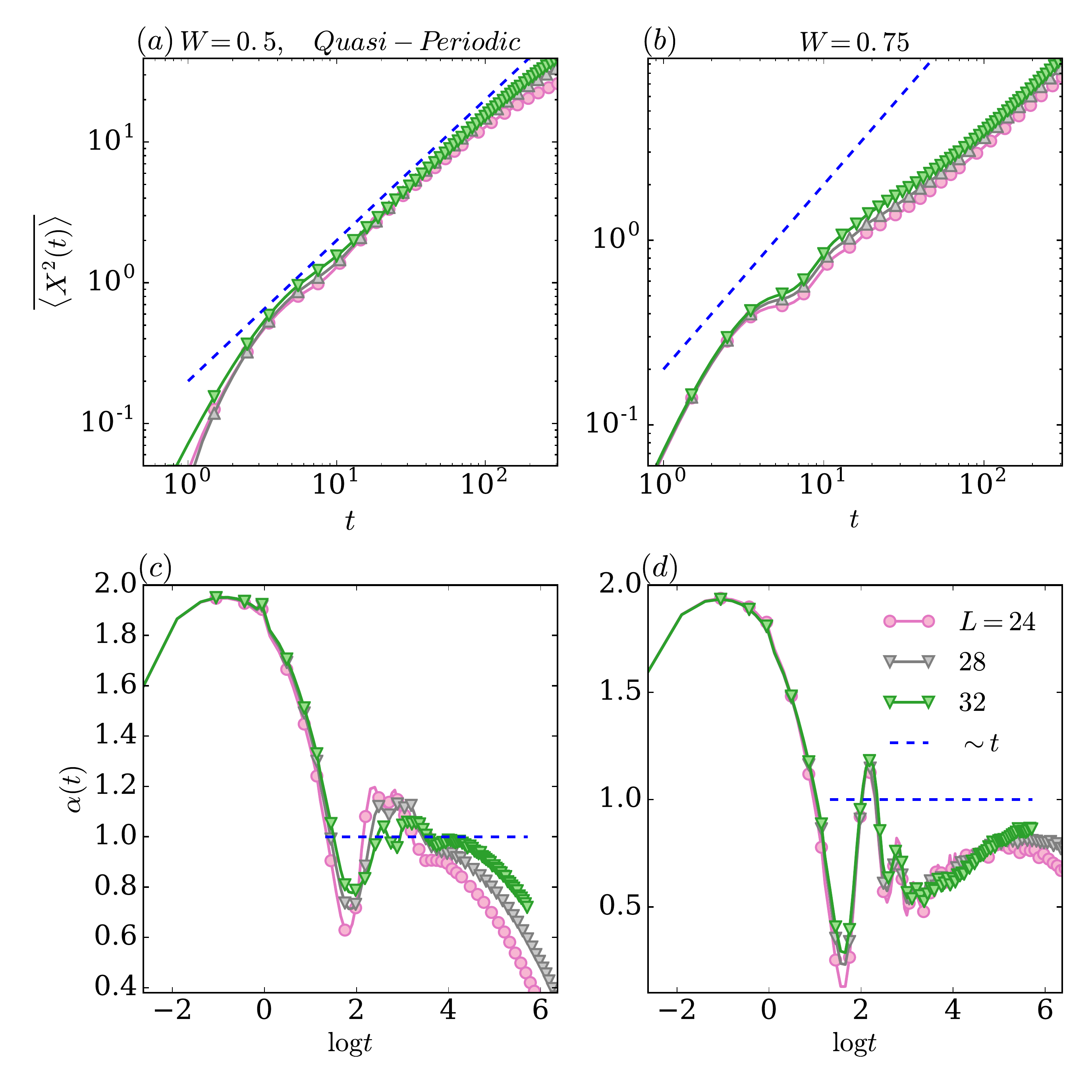}
\caption{(a): Averaged width $\overline{ \langle X^2(t)\rangle }$ of the density propagator $\Pi(x,t)$ at fix $W=0.5$ for several $L\in \{24,28,32\}$.  (b) Dynamical exponent $\alpha(t) = {d \log{\overline{\langle X^2(t) \rangle} }}/{d \log{t}}$ at $W=0.5$, with the enlarging plateau close to $\alpha=1$ (diffusion).  The dashed line in (a) and (b) is a guide for eyes and represents the diffusive behavior $\sim t$.  (c) $\overline{ \langle X^2(t)\rangle }$ at $W=0.75$ (value close to the MBL transition $W/W_c\approx 0.42$).  (d) $\alpha(t)$ at $W=0.75$ for several $L$.  In this case $\alpha(t)$ does not form a plateau at large time, instead  it increases and might approach $\alpha =1$ (dashed line) in the thermodynamic limit.}
\label{fig:Fig3_QP}
\end{figure} 

In what follows we investigate the out-of-equilibrium dynamics of $\hat H_\infty$ restricted to the block containing $N_{\bullet\bullet}-1$ movers. 

Figure~\ref{fig:Fig3_QP}~(a) shows $\langle X^2(t) \rangle$ as a function of time for a fixed $W = 0.5$ and several $L\in \{24, 28, 32\}$. 

At short times the propagation is ballistic $\langle X^2(t) \rangle\sim t^2 $.
Instead, at longer time scales, transient dynamics are observed which culminate in a diffusive propagation $\langle X^2(t) \rangle \sim t$ 
(dashed line in Fig.~\ref{fig:Fig3_QP}~(a)). To better pin down the behavior of $\langle X^2(t) \rangle$ 
we study its dynamical exponent $\alpha(t)$ in Eq.~\ref{eq:dynamical_exp}. Fig.~\ref{fig:Fig3_QP}~(c) shows $\alpha(t)$ for $W=0.5$. 
After the ballistic propagation $\alpha(t)\approx 2$ and transient dynamics, $\alpha(t)$ forms a plateau around $\alpha =1$ (dashed line in Fig.~\ref{fig:Fig3_QP}~(c)),
giving thus indication that the propagation is diffusive. 

For values closer to the MBL transition we found the same transient dynamics that we report for the case with random potential.
This transient dynamics seems to approach a diffusive propagation, as shown in Fig.~\ref{fig:Fig3_QP}~(c) -- (d). 
Figure~\ref{fig:Fig3_QP}~(d) shows $\alpha(t)$ for $W = 0.75$ ($W/W_c\approx 0.47$). Close to the MBL transition within its ergodic phase, 
$\alpha(t)$ does still increases with time and it might reach the diffusive value $\alpha = 1$ (dashed line in Fig.~\ref{fig:Fig3_QP}~(d)) at larger times. 

 \bibliography{V_inf_bib}

\end{document}